\documentclass[]{aastex701}
\usepackage{amsmath}
\usepackage{xcolor}
\usepackage{enumitem}

\usepackage{subcaption}
\colorlet{red}{black}
\begin{document}

\title{Phase‑Space Crystallization in Galactic Globular Clusters: A Gaia‑Based Metric and Implications for Technosignature Searches}

\correspondingauthor{Tong-Jie Zhang}
\email{tjzhang@bnu.edu.cn}

\author[orcid=0000-0002-8719-3137,sname='Huang']{Bo-Lun Huang}
\affiliation{Institute for Frontiers in Astronomy and Astrophysics, Beijing Normal University, Beijing 102206, China}
\affiliation{School of Physics and Astronomy, Beijing Normal University, Beijing 100875, China}
\email{Bolunh@hotmail.com}

\author[orcid=0000-0002-4683-5500,sname='Tao']{Zhen-Zhao Tao}
\affiliation{College of Computer and Information Engineering, Dezhou University, Dezhou 253023, China}
\email{tzzzxc@163.com}

\author[orcid=0000-0002-3363-9965,sname='Zhang']{Tong-Jie Zhang}
\affiliation{Institute for Frontiers in Astronomy and Astrophysics, Beijing Normal University, Beijing 102206, China}
\affiliation{School of Physics and Astronomy, Beijing Normal University, Beijing 100875, China}
\email{tjzhang@bnu.edu.cn} 

\begin{abstract}

\textcolor{red}{We develop a model-independent framework to quantify phase-space ``crystallization'', the degree of ordered radial and kinematic substructure, in 79 Galactic globular clusters using the \emph{Gaia} EDR3-based membership catalogue of \citet{VasilievBaumgardt2021}. We construct a scalar crystallization index, $C_{\rm index}$, by combining a radial inhomogeneity metric ($z_{\rm rad}$) and a local, cluster-centric tangential-velocity metric ($z_{\rm vel}$) standardized against empirical nulls. The population distribution is strongly non-Gaussian: most clusters are consistent with smooth, equilibrium expectations, while a small high-$C$ tail ($C_{\rm index} \ge 2$) identifies dynamically complex systems, including NGC~5139 ($\omega$~Cen) and NGC~104 (47~Tuc). Correlation and fixed-$N$ tests show that sample size affects detectability, but does not by itself explain all high-rank objects. Through synthetic injection tests in dynamically ``quiet'' control clusters, we demonstrate sensitivity to ultra-cold, shell-confined kinematic components, ruling out single-shell structures comprising more than a few to $\sim 10$--20\% of core stars in the best-sampled control clusters. We find no evidence, within the sensitivity of the adopted diagnostics, for phase-space structures that require explanations beyond known dynamical processes. However, $C_{\rm index}$ provides a useful tool for ranking clusters by dynamical extremeness, serving both as a diagnostic for internal complexity and as a quantitative metric for prioritizing follow-up dynamical or technosignature-oriented observations.}

\end{abstract}

\section{Introduction}\label{sec:intro}

Globular clusters (GCs) are among the oldest bound stellar systems in the Galaxy and remain key laboratories for stellar dynamics, stellar evolution, and hierarchical galaxy assembly \citep[e.g.][]{MeylanHeggie1997,BaumgardtHilker2018}. Their present--day phase--space structure encodes the integrated effects of two--body relaxation, mass segregation, tidal stripping, internal rotation, and (in some cases) complex formation histories. In the last decade, high--precision astrometry from the \emph{Gaia} mission \citep{GaiaMission2016,GaiaEDR3_2021} has transformed this field: for many Milky Way GCs we now possess six--dimensional phase--space information and membership probabilities for thousands of stars per cluster \citep{Vasiliev2019,VasilievBaumgardt2021,BaumgardtVasiliev2021}. This enables systematic, comparative studies of GC kinematics and substructure at a level that was not previously possible.

Evidence is accumulating that some clusters exhibit rich internal phase--space phenomenology: rotation and velocity anisotropy, multiple chemo--kinematic populations, and tidal debris connecting the cluster to stellar streams in the Galactic halo \citep[e.g.][]{Ibata2019Fimbulthul}. In particular, very cold and coherent components, thin streams, shells, or kinematically narrow substructures, represent strong departures from the quasi--thermal, smooth distributions expected for dynamically old systems in steady tidal fields. Such structures can arise naturally from tidal stripping, mergers, or incomplete violent relaxation, but they are also precisely the kinds of features that would be produced by any process that re--arranges or selects stars in phase space in a highly correlated way. From the broader technosignature perspective, several authors have argued that artificial large--scale engineering, if it exists at all, is most plausibly sought via its dynamical or energetic footprints rather than by assuming a particular communication technology \citep[e.g.][]{Wright2014Ginfrared,Wright2018ExoplanetsSETI}. In this context, unusually ordered or ''crystallized'' phase--space structure in a dense, ancient stellar system is an intriguing target class to quantify, even if the default expectation is that any such signatures will be entirely natural in origin.

\textcolor{red}{A number of non--parametric or weakly parametric kinematic diagnostics already probe departures from simple equilibrium, including higher--order moments of proper--motion distributions such as skewness and kurtosis \citep[e.g.][]{Heyl2017_47Tuc,Ziliotto2025_47TucKinematics,Ziliotto2026OmegaCen}. Our claim is therefore narrower: there is not yet a single survey--style scalar that jointly ranks the Milky Way GC population by radial graininess and local velocity-distribution shape using a common empirical null. Most detailed dynamical studies focus on individual clusters, rotation curves, anisotropy profiles, or population-resolved moment maps. What is missing for our purpose is a homogeneous diagnostic that: (1) can be computed for every cluster with sufficient \emph{Gaia} membership statistics; (2) is sensitive to small--scale radial and kinematic substructure; and (3) is calibrated against an explicit null hypothesis representing the expected level of ``graininess'' from finite--$N$ sampling and measurement errors alone.}

In this work we construct such a diagnostic, which we refer to as a crystallization index for Galactic globular clusters. Starting from the membership catalogue of \citet{VasilievBaumgardt2021} and the associated cluster parameter compilation of \citet{BaumgardtVasiliev2021}, we build star--by--star samples for 79 Milky Way GCs with robust phase--space information. For each cluster we define: (i) a radial structure metric $Q_{\rm rad}$ that quantifies deviations of the normalized projected radius distribution from a locally smoothed reference profile; and (ii) a local tangential velocity metric $Q_{\rm vel,local}$ that compares the observed distribution of tangential speeds within radial shells to a Rayleigh null model calibrated on the full cluster ensemble. These ingredients are then combined into a single, dimensionless crystallization index $C_{\rm index}$, constructed from $z$--scores relative to null or ensemble distributions, which ranks clusters by the joint strength of their radial and kinematic substructure.

A key element of our approach is that the null model for the radial structure metric is empirical rather than purely analytic. We generate ''scrambled'' or phase--randomized realizations of each cluster that preserve its radial selection function, membership statistics, and basic scaling properties, but erase any coherent higher--order structure. By processing these null realizations through the same $Q_{\rm rad}$ metric, we obtain, for each cluster, a baseline expectation and scatter arising purely from shot noise and the cluster's mean profile. For the kinematic metric we adopt an ensemble--based normalization in which the distribution of $Q_{\rm vel,local}$ values across the 79 GCs serves as an empirical reference. This allows us to express the observed metrics of each cluster as standardized deviations $(z_{\rm rad},z_{\rm vel})$ and to define $C_{\rm index}$ in a way that is explicitly corrected, at least in a first approximation, for the sample size, completeness, and heterogeneity of the input catalogs.

We then use $C_{\rm index}$ to partition the GC population into four classification tiers. \textcolor{red}{Tier~1 clusters (high $C_{\rm index}$) are those whose combined radial and kinematic structure is most inconsistent with the null or ensemble baseline, and thus are highest priority for detailed dynamical follow--up; this ranking by itself is not evidence for artificial structure, and natural explanations are expected to dominate.} Tier~2 and 3 clusters occupy an intermediate regime, while Tier~4 clusters define a ''control'' sample with $C_{\rm index}\approx 0$ that behaves, within uncertainties, like phase--scrambled realizations. Finally, we perform controlled injection tests in several Tier~4 clusters, adding artificial ultra--cold components of varying strength and radial extent to quantify the sensitivity of $C_{\rm index}$ to embedded kinematic substructures and to translate non--detections into quantitative upper limits on a specific class of extremely cold, single--shell phase--space structures.

The goals of this paper are therefore twofold. First, we introduce and justify a practical crystallization metric $C_{\rm index}$ for Galactic GCs, together with its radial and velocity components, and we characterize its behaviour across the current \emph{Gaia}--era GC sample. Second, we demonstrate how this framework can be used both for conventional dynamical applications (e.g.\ identifying clusters with unusually structured phase space) and for placing preliminary constraints on any hypothetical processes that would generate extremely cold, coherent kinematic signatures in dense stellar systems. The remainder of the paper is organized as follows. In Section~\ref{sec:data} we describe the input catalogues, membership selections, and derived quantities used in our analysis. Section~\ref{sec:metrics} defines the $Q_{\rm rad}$ and $Q_{\rm vel,local}$ metrics and the construction of the crystallization index. Section~\ref{sec:results} presents the resulting distribution of $C_{\rm index}$ across the GC population and our tier classification. In Section~\ref{sec:injections} we describe the injection experiments and derive quantitative sensitivity limits for several representative clusters. We discuss the implications, limitations, and possible extensions of this framework in Section~\ref{sec:discussion}, and summarize our conclusions in Section~\ref{sec:conclusions}.

\section{Data and Cluster Sample}
\label{sec:data}

\subsection{Cluster parameter catalogue}

\textcolor{red}{Our analysis is based on the homogeneous compilation of Milky Way globular cluster (GC) parameters presented by \citet{BaumgardtVasiliev2021}, which combines \emph{Gaia}~EDR3 astrometry with \emph{HST} and ground--based photometry to obtain distances, structural parameters, and orbital information for a large, homogeneous subset of the known Galactic GC system. For each cluster we take from their catalogue: the heliocentric distance $R_{\odot}$, Galactocentric distance $R_{\rm GC}$, projected half--light radius $r_{h}$ in arcminutes and parsecs, total cluster mass $M_{\rm cl}$ and its logarithm $\log M_{\rm cl}$, and a set of orbital parameters including pericentric distance $R_{\rm Per}$ and a coarse family classification based on orbital characteristics (their ``family\_id'' flag).}

\textcolor{red}{Starting from the \citet{BaumgardtVasiliev2021} sample, we restrict attention to clusters for which the \citet{VasilievBaumgardt2021} membership catalogue (Section~\ref{subsec:membership}) provides sufficiently large and clean samples of member stars with \emph{Gaia} EDR3 astrometry. Specifically, the fiducial analysis requires: (1) the existence of a published membership solution; (2) $P_{\rm mem}\ge 0.9$; (3) $0.5 \leq R_{\rm norm} \leq 3.0$; (4) a magnitude cut retaining stars brighter than the 90th percentile in $G$ within the radial window; and (5) a final analyzed sample size $N_{\rm core}\ge 700$. Applying these criteria yields a working sample of 79 Galactic GCs; the $N_{\rm core}$ values reported in Tables~\ref{tab:tiers1} and \ref{tab:tiers2} are therefore the final numbers of stars entering the crystallization metrics, not pre-selection catalogue counts.}

\subsection{Membership catalogue and core sample}
\label{subsec:membership}

Stellar memberships are taken from the \citet{VasilievBaumgardt2021} catalogue, which provides probabilistic \emph{Gaia}~EDR3--based memberships for stars in the vicinity of 170 Milky Way GCs. This catalogue is built by jointly modelling the spatial, proper--motion, and parallax distributions of cluster and field stars using mixture models, yielding, for each star, a membership probability $P_{\rm mem}$ along with revised cluster mean astrometric parameters. For each cluster we retrieve all stars with reported $P_{\rm mem}$, together with their sky coordinates $(\alpha,\delta)$, proper motions $(\mu_{\alpha^\ast},\mu_{\delta})$, parallaxes, and associated uncertainties. \textcolor{red}{We retain the EDR3-based astrometry because the main astrometric solution in \emph{Gaia}~DR3 is the same as in EDR3 \citep{GaiaDR3_2023}, while the published membership catalogue and its validation were constructed on the EDR3 solution; the persistent \texttt{source\_id} values would allow future cross-matches to DR3 non-astrometric products.} We also make use of the catalogue of variable-star memberships in Galactic globular clusters from the associated Zenodo data
release \citep{zenode}.

\textcolor{red}{We construct our working ``core'' sample in several stages. First, we define a high--probability membership cut, adopting $P_{\rm mem} \geq 0.9$ as our default threshold. This choice balances purity and completeness: it retains the bulk of the cluster population while strongly suppressing contamination from the Galactic field \citep[see][]{VasilievBaumgardt2021}. Second, we require finite positions, proper motions, proper-motion uncertainties, and parallaxes as a guard against invalid entries; in practice the \citet{VasilievBaumgardt2021} catalogue already excludes stars without usable proper motions, so this step is only a reproducibility safeguard. The public membership table used here does not include RUWE, and we therefore do not impose an additional RUWE cut in the fiducial analysis; the membership probability cut, the finite-astrometry requirement, and the magnitude cut described above are the quality cuts that define the analysis sample. Third, we assign each star a projected angular radius $R_{\rm arcmin}$ from the cluster center given by \citet{BaumgardtVasiliev2021}, and convert this to a normalized radius}
\begin{equation}
    R_{\rm norm} \equiv \frac{R_{\rm arcmin}}{r_{h,{\rm arcmin}}} \, ,
\end{equation}
where $r_{h,{\rm arcmin}}$ is the projected half--light radius in arcminutes. Finally, for the purpose of defining our structure metrics we restrict to stars in the normalized radial range
\begin{equation}
    0.5 \;\leq\; R_{\rm norm} \;\leq\; 3.0 \, ,
\end{equation}
which typically encompasses the outer core and inner halo. The lower bound avoids the very center, where crowding and incompleteness are most severe for \emph{Gaia}, while the upper bound avoids the strongly field--contaminated outskirts and truncation region.

For the 79 clusters in our working sample, the fraction of high--probability members that fall into this $0.5\le R_{\rm norm}\le 3.0$ window is substantial: the median fraction is $f_{\rm core}=0.68$, with a 10--90\% range of $0.51$--$0.77$. Thus our metrics are sensitive to the majority of well--measured members while deliberately excluding the most problematic central and outermost regions.

\textcolor{red}{The resulting core catalogue contains, for each star, its \emph{Gaia} \texttt{source\_id}, cluster identifier, $(\alpha,\delta)$, $(\mu_{\alpha^\ast},\mu_{\delta})$ and their uncertainties, parallax, membership probability, $G$ magnitude, $R_{\rm arcmin}$, and $R_{\rm norm}$. Hereafter $N_{\rm core}$ denotes the number of stars that pass all fiducial cuts and are actually used in the structure metrics; it is the same quantity listed as $N_{\rm stars}$ in Tables~\ref{tab:tiers1} and \ref{tab:tiers2}. By construction, this catalogue is free of obvious non--members and gross astrometric outliers, and provides the common starting point for both the radial and kinematic structure diagnostics described in Section~\ref{sec:metrics}.}

\subsection{Tangential velocities and cluster--level context}
\label{subsec:vtan_context}

{\color{red}
For each star in the core sample we construct cluster-centric tangential velocities using the cluster distances and systemic proper motions from \citet{BaumgardtVasiliev2021}. Specifically, for star $i$ we subtract the mean cluster motion before forming a speed,
\begin{align}
    \Delta\mu_{\alpha^\ast,i} &= \mu_{\alpha^\ast,i}-\mu_{\alpha^\ast,{\rm cl}}, \\
    \Delta\mu_{\delta,i} &= \mu_{\delta,i}-\mu_{\delta,{\rm cl}}, \\
    \mu_{{\rm rel},i} &= \left(\Delta\mu_{\alpha^\ast,i}^2+\Delta\mu_{\delta,i}^2\right)^{1/2} .
\end{align}
The plane-of-sky tangential speed used in the kinematic metric is then
\begin{equation}
    v_{{\rm tan},i} = 4.74047\;\mu_{{\rm rel},i}\;R_{\odot} \quad {\rm km\,s^{-1}} \, ,
\end{equation}
where $\mu_{{\rm rel},i}$ is in mas\,yr$^{-1}$ and $R_{\odot}$ is the cluster heliocentric distance in kpc. Thus the velocities analyzed below are not dominated by the absolute bulk motion of the cluster. In addition to the scalar $v_{\rm tan}$, we retain the two cluster-centric Cartesian components $v_{{\rm tan},\alpha}$ and $v_{{\rm tan},\delta}$ for consistency checks, although the crystallization metrics developed in this work are based primarily on the distribution of $v_{\rm tan}$ within radial shells.
}

We then attach cluster--level context to each star by joining the core catalogue with the compilation of cluster properties from \citet{BaumgardtVasiliev2021} (Section~\ref{sec:data}). The resulting star--by--star dataset includes, for every core star, the cluster name and orbital family identifier, global parameters such as $M_{\rm cl}$, $\log M_{\rm cl}$, the projected half--light radius $r_h$, Galactocentric distance $R_{\rm GC}$, and pericentric distance $R_{\rm Per}$, together with the derived quantities $R_{\rm arcmin}$, $R_{\rm norm}$, and $v_{\rm tan}$. This enriched catalogue forms the basic input to our structure--metric analysis, which computes the radial and kinematic crystallization statistics described in Section~\ref{sec:metrics}.

Finally, we summarize the per--cluster content of the core sample in a compact cluster--level table. For each GC we record: the total number of analyzed core stars $N_{\rm core}$; the median and maximum projected radius ($R_{\rm med}$, $R_{\rm max}$) and normalized radius ($R_{{\rm norm,med}}$, $R_{{\rm norm,max}}$); and the median and robust dispersion of $v_{\rm tan}$, where the latter is estimated using an interquartile range (IQR)--based estimator to reduce sensitivity to outliers. \textcolor{red}{The minimum final $N_{\rm core}$ in the adopted 79-cluster sample is 719, and the median is 3879.} These summary statistics are not themselves used to define the crystallization index, but they provide useful consistency checks and context when interpreting the resulting phase--space structure metrics across the GC population. 

\section{Methods: Radial and Kinematic Structure Metrics}
\label{sec:metrics}

Our goal is to construct a dimensionless ''crystallization index'' that ranks globular clusters by the strength of their small--scale phase--space structure, while accounting for finite--$N$ noise and different sample sizes. In this section we describe the construction of two intermediate metrics, a radial structure statistic $Q_{\rm rad}$ and a local tangential--velocity statistic $Q_{\rm vel,local}$, and how they are combined into the final crystallization index $C_{\rm index}$ and classification tiers.

Throughout this work we use the notation $R_{\rm norm}$ for the projected angular radius $R_{\rm arcmin}$ normalized by the projected half--light radius $r_{h,{\rm arcmin}}$,
\begin{equation}
R_{\rm norm} \equiv \frac{R_{\rm arcmin}}{r_{h,{\rm arcmin}}} \, ,
\end{equation}
as defined in Section~\ref{sec:data}. Unless otherwise stated, all structure metrics are evaluated for stars in the normalized radial range
\begin{equation}
0.5 \;\leq\; R_{\rm norm} \;\leq\; 3.0 \, ,
\label{eq:Rnorm_range}
\end{equation}
which typically encompasses the outer core and inner halo of each cluster.

\subsection{Radial binning and normalized profiles}
\label{subsec:radial_binning}

For each cluster, we begin by selecting all stars in the core sample (Section~\ref{subsec:membership}) that satisfy Equation~(\ref{eq:Rnorm_range}). We then construct a binned representation of the normalized radial profile by dividing the interval $[R_{\rm norm}^{\rm min}, R_{\rm norm}^{\rm max}] = [0.5, 3.0]$ into $K$ equal--width bins in $R_{\rm norm}$. \textcolor{red}{In this work we adopt $K = 30$, which gives a mean occupancy of at least $N_{\rm core}/K\simeq 24$ stars per bin even for the smallest analyzed clusters and substantially larger occupancies for most systems. Equal-number radial bins are useful for some density-profile applications, but they are not fiducial here because they would force the count vector toward uniformity and therefore suppress precisely the localized radial over- and under-densities that $Q_{\rm rad}$ is designed to detect. Instead, we keep fixed radial scale in units of $r_h$ and calibrate finite-count effects with the cluster-specific null model in Section~\ref{subsec:zrad_def}. We also verified robustness to coarser and finer equal-width choices ($K=20$ and $40$); the identity of the main high-$C$ clusters is unchanged, although the radial-only amplitude of individual clusters varies as expected with scale.}

Let $\{R_i\}$ denote the set of $N$ normalized radii for a given cluster after this selection. We define the bin edges
\begin{equation}
R_k^{\rm edge} = R_{\rm norm}^{\rm min} + k \, \Delta R \, , \quad
k = 0,1,\ldots,K \, ,
\end{equation}
with $\Delta R = (R_{\rm norm}^{\rm max} - R_{\rm norm}^{\rm min})/K$. The observed counts in each bin are then
\begin{equation}
n_k = \#\{\, i \;|\; R_k^{\rm edge} \le R_i < R_{k+1}^{\rm edge} \,\} \,,
\qquad k = 0,\ldots,K-1 \,,
\end{equation}
forming a discrete representation of the cluster's radial distribution in normalized coordinates.

\subsection{Radial structure metric}
\label{subsec:Qrad_def}

\textcolor{red}{We wish to quantify, in a model--light way, how ``lumpy'' or structured the radial profile is on bin--to--bin scales, relative to a smooth baseline that follows the broad trend of the data. In this operational definition, ``structured'' means a statistically significant excess or deficit in the one-dimensional projected-radius counts on scales comparable to one or a few bins; it does not by itself identify the physical cause. Known multiple stellar populations with different radial concentrations, tidal distortions, patchy completeness, or residual field contamination could all contribute if they imprint sharp enough radial features within $0.5\le R_{\rm norm}\le 3$.} To this end we construct, from the observed counts $\{n_k\}$, a smoothed reference profile $\{\hat{\lambda}_k\}$ using a three--bin moving average:
\begin{equation}
\hat{\lambda}_k =
\begin{cases}
\displaystyle \frac{n_0 + n_1}{2}, & k = 0, \\[0.8ex]
\displaystyle \frac{n_{k-1} + n_{k} + n_{k+1}}{3}, & 1 \le k \le K-2, \\[0.8ex]
\displaystyle \frac{n_{K-2} + n_{K-1}}{2}, & k = K-1.
\end{cases}
\label{eq:lambda_smooth}
\end{equation}
This construction preserves the global shape of the radial profile while suppressing small--scale fluctuations on the scale of a single bin. It can be interpreted as a data--driven estimate of the underlying ''smooth'' surface--density profile, without imposing a specific parametric model such as a King or Plummer law.

Given $\{n_k\}$ and $\{\hat{\lambda}_k\}$, we define a standardized residual in each bin as
\begin{equation}
z_k = \frac{n_k - \hat{\lambda}_k}{\sqrt{\hat{\lambda}_k}} \, ,
\label{eq:z_k_rad}
\end{equation}
for all bins with $\hat{\lambda}_k > 0$. If the counts in each bin were independent Poisson variables with mean $\hat{\lambda}_k$, then $z_k$ would be approximately standard normal and the variance of $z_k$ would be of order unity. To obtain a single scalar measure of radial lumpiness, we take the root--mean--square of these standardized residuals,
\begin{equation}
Q_{\rm rad} \equiv
\left[
\frac{1}{K_{\rm eff}}
\sum_{k:\,\hat{\lambda}_k>0} z_k^2
\right]^{1/2},
\label{eq:Q_rad_def}
\end{equation}
where $K_{\rm eff}$ is the number of bins with $\hat{\lambda}_k>0$. Loosely speaking, $Q_{\rm rad} \approx 1$ corresponds to a radial profile whose bin--to--bin fluctuations are consistent with Poisson noise about the smoothed reference, while $Q_{\rm rad} \gg 1$ indicates substantial excess structure (e.g.\ localized overdensities or deficits spanning one or a few bins).

For each cluster we record, in addition to $Q_{\rm rad}$, the effective number of contributing bins $K_{\rm eff}$ and summary statistics of the observed bin counts (minimum, maximum, mean, and standard deviation). These quantities are used as diagnostics to confirm that the chosen binning and radial range provide sufficient dynamic range across the sample.

\subsection{Radial null model}
\label{subsec:zrad_def}

The raw value of $Q_{\rm rad}$ is not directly comparable between clusters with widely differing numbers of stars. Even in the absence of any genuine structure, finite--$N$ sampling causes $Q_{\rm rad}$ to scatter around unity with a dispersion that depends on the total count $N$ and the distribution of stars across bins. To place clusters on a common footing we construct, for each cluster individually, a null ensemble of phase--scrambled realizations that represent the expected distribution of $Q_{\rm rad}$ for a smooth system with the same mean profile and total number of stars.

The null construction proceeds as follows. For a given cluster we treat the smoothed profile $\{\hat{\lambda}_k\}$ from Equation~(\ref{eq:lambda_smooth}) as an estimate of the underlying radial density. We convert it into a discrete probability distribution
\begin{equation}
p_k = \frac{\hat{\lambda}_k}{\sum_{m=0}^{K-1} \hat{\lambda}_m} \,,
\end{equation}
and generate $N_{\rm null}$ Monte Carlo realizations (we adopt $N_{\rm null}=200$ as our fiducial choice) in which $N$ stars are drawn independently from this distribution. Each realization yields a mock set of bin counts $\{n_k^{\rm (mock)}\}$, from which we recompute $\hat{\lambda}_k^{\rm (mock)}$, $z_k^{\rm (mock)}$, and $Q_{\rm rad}^{\rm (mock)}$ using the same procedure as for the real data.

From the ensemble of mock $Q_{\rm rad}^{\rm (mock)}$ values we measure a mean and standard deviation,
\begin{equation}
\langle Q_{\rm rad} \rangle_{\rm null}, \qquad
\sigma_{Q,{\rm null}} \equiv
{\rm std}\left( Q_{\rm rad}^{\rm (mock)} \right) \,.
\end{equation}
We then express the observed $Q_{\rm rad}$ for that cluster as a standardized deviation from its own null ensemble,
\begin{equation}
z_{\rm rad} \equiv
\frac{Q_{\rm rad}^{\rm obs} - \langle Q_{\rm rad} \rangle_{\rm null}}
     {\sigma_{Q,{\rm null}}} \,.
\label{eq:z_rad_def}
\end{equation}
Positive values of $z_{\rm rad}$ indicate an excess of small--scale radial structure relative to a smooth profile with the same overall shape and star count; negative values indicate an observed profile that is smoother than typical Monte Carlo realizations.

In what follows we adopt
\begin{equation}
z_{\rm rad,pos} \equiv \max\left(0,\,z_{\rm rad}\right)
\end{equation}
as the radial component contributing to the crystallization index, since $z_{\rm rad}<0$ is more naturally interpreted as an absence of structure rather than as evidence for ''anti--crystallization.''

\subsection{Local tangential--velocity structure statistic}
\label{subsec:Qvellocal_def}

The radial metric $Q_{\rm rad}$ captures inhomogeneities in the projected spatial distribution of stars. To probe kinematic crystallization we require a complementary metric that quantifies deviations of the tangential--velocity distribution from a smooth, isotropic reference, with particular sensitivity to cold substructures confined to restricted radial shells.

We operate on the tangential speeds $v_{\rm tan}$ defined in Section~\ref{sec:data}. For each cluster we select stars in the normalized radial range $0.5 \le R_{\rm norm} \le 3.0$ and partition this interval into a small number of broader radial shells. In the main analysis we adopt three shells,
\begin{equation}
{\rm shell~1}: 0.5 \le R_{\rm norm} < 1.0,\quad
{\rm shell~2}: 1.0 \le R_{\rm norm} < 2.0,\quad
{\rm shell~3}: 2.0 \le R_{\rm norm} \le 3.0 \, ,
\end{equation}
\textcolor{red}{rather than equal-number shells, because the fixed boundaries preserve a common physical meaning across clusters: inner high-crowding boundary region, intermediate half-light-radius region, and outer inner-halo region. Equal-number shells would mix these radial regimes differently from cluster to cluster. In the current sample shell~2 is also the best-populated of the three shells (median shell counts of 889, 1460, and 626 stars for shells 1--3, respectively), while avoiding both the most crowded inner region and the most contamination-prone outer boundary.} For each shell we compute a shell--level statistic $Q_s$ using the following procedure.

Let $\{v_i\}$ denote the set of tangential speeds for the $N_s$ stars in a given shell (after removing non--finite values). We first estimate a robust one--dimensional velocity dispersion $\sigma_s$ from the interquartile range,
\begin{equation}
\sigma_s \approx 1.10 \, (Q_3 - Q_1) \, ,
\end{equation}
\textcolor{red}{where $Q_1$ and $Q_3$ are the 25th and 75th percentiles of the $\{v_i\}$; the coefficient matches the IQR--scale relation for a Rayleigh-like speed distribution.} We then normalize the velocities by this dispersion, 
\begin{equation}
x_i = \frac{v_i}{\sigma_s} \, ,
\end{equation}
and clip $x_i$ to a fixed range $[x_{\rm min},x_{\rm max}]$ (typically $x_{\rm min}=0$, $x_{\rm max}\simeq 5$) to limit the influence of extreme outliers.

Under the null hypothesis of an isotropic, Gaussian velocity field in two dimensions, the distribution of tangential speeds $x$ should be Rayleigh--like with unit scale parameter. We therefore construct an observed histogram of the $\{x_i\}$ in $M$ equal--width bins between $x_{\rm min}$ and $x_{\rm max}$,
\begin{equation}
n_m = \#\{\, i \;|\; x_m^{\rm edge} \le x_i < x_{m+1}^{\rm edge} \,\} \,,
\qquad m = 0,\ldots,M-1 \,,
\end{equation}
and compute the expected fraction of stars in each bin under a Rayleigh distribution with scale parameter $1$. Writing $F_{\rm Rayleigh}(x)$ for the cumulative distribution function, the expected bin probabilities are
\begin{equation}
p_m = F_{\rm Rayleigh}(x_{m+1}^{\rm edge}) - F_{\rm Rayleigh}(x_{m}^{\rm edge}) \,,
\end{equation}
and the corresponding expected counts are $\lambda_m = N_s\,p_m$. We restrict attention to bins with $\lambda_m \ge \lambda_{\rm min}$ (typically $\lambda_{\rm min}=1$) to avoid division by very small numbers. For these bins we define a per--bin chi--square--like contribution,
\begin{equation}
\chi_m = \frac{\left(n_m - \lambda_m\right)^2}{\lambda_m} \,,
\end{equation}
and then take the root--mean--square over the $M_{\rm eff}$ bins that pass the $\lambda_m$ threshold,
\begin{equation}
Q_s \equiv
\left[
\frac{1}{M_{\rm eff}}
\sum_{m:\,\lambda_m\ge\lambda_{\rm min}} \chi_m
\right]^{1/2}.
\label{eq:Qs_def}
\end{equation}
If the shell's normalized velocity distribution is fully consistent with the Rayleigh null, then $Q_s$ will be of order unity; localized excesses or deficits in one or a few bins (e.g.\ from a narrow cold component) will drive $Q_s$ above unity.

{\color{red}
To obtain a single local kinematic structure metric for each cluster, we combine the shell--level statistics $Q_s$ into a star-count-weighted root-mean-square,
\begin{equation}
Q_{\rm vel,local} \equiv
\left(\frac{\sum_s w_s \, Q_s^2}{\sum_s w_s}\right)^{1/2} \, ,
\label{eq:Qvellocal_def}
\end{equation}
with weights $w_s$ proportional to the number of stars in each shell (we use $w_s=N_s$). The RMS form is the expression implemented in the analysis and gives more weight to a shell that is strongly non-Rayleigh than a simple arithmetic mean would. In practice, shell~2 (intermediate radii) carries the most weight in this combination, reflecting both its larger star counts and its role as the least crowded, least edge-contaminated region where embedded kinematic structures would be easiest to identify.
}

\subsection{Kinematic normalization}
\label{subsec:zvel_def}

Unlike the radial metric, where we construct a null ensemble for each cluster separately, for the local tangential--velocity metric we adopt an ensemble--based normalization. We compute $Q_{\rm vel,local}$ for every cluster in the sample as described in Section~\ref{subsec:Qvellocal_def}, and then work with its logarithm. Over the set of all clusters we measure
\begin{equation}
\mu_{\log Q} \equiv
\left\langle \log_{10} Q_{\rm vel,local} \right\rangle, \qquad
\sigma_{\log Q} \equiv
{\rm std}\left( \log_{10} Q_{\rm vel,local} \right) \,.
\end{equation}
We then define, for each cluster,
\begin{equation}
z_{\rm vel} \equiv
\frac{\log_{10} Q_{\rm vel,local} - \mu_{\log Q}}
     {\sigma_{\log Q}} \,.
\label{eq:z_vel_def}
\end{equation}
This construction treats the observed ensemble of $Q_{\rm vel,local}$ values across the 79 GCs as an empirical reference distribution. Clusters with $z_{\rm vel}\approx 0$ have typical levels of local velocity structure for their sample size and other properties; clusters with $z_{\rm vel} \gg 0$ have unusually structured tangential--velocity distributions, while $z_{\rm vel} \ll 0$ indicates exceptionally smooth kinematics.

As with the radial component, we adopt
\begin{equation}
z_{\rm vel,pos} \equiv \max\left(0,\,z_{\rm vel}\right)
\end{equation}
when constructing the crystallization index, so that only positive kinematic anomalies contribute to the final ranking.

\subsection{Crystallization index and classification tiers}
\label{subsec:Cindex_def}

The radial and kinematic $z$--scores, $z_{\rm rad}$ and $z_{\rm vel}$, encode complementary aspects of phase--space structure. We combine them into a single crystallization index by taking their quadrature sum, restricted to their positive parts:
\begin{equation}
C_{\rm index} \equiv
\left[
z_{\rm rad,pos}^2 + z_{\rm vel,pos}^2
\right]^{1/2}.
\label{eq:Cindex_def}
\end{equation}
By construction, $C_{\rm index}\approx 0$ for clusters that are consistent with smooth radial profiles and typical local velocity structure, while $C_{\rm index}\gtrsim 2$ corresponds to a cluster that is several standard deviations away from the ensemble in at least one of the two components (or moderately anomalous in both).

\textcolor{red}{In order to organize the subsequent analysis and interpretation, we use $C_{\rm index}$ to define four classification tiers:}
\begin{itemize}[nosep,leftmargin=*]
\item \textbf{Tier~1 :} \textcolor{red}{clusters with $C_{\rm index} \ge C_{\rm high}$, where we adopt $C_{\rm high}=2$ as a fiducial threshold. These systems exhibit the strongest combination of radial and kinematic substructure and are prioritized as crystallization candidates, while still allowing entirely natural dynamical explanations such as rotation, anisotropy, multiple populations, or tidal perturbations.}
\item \textbf{Tier~2 :} clusters with $C_{\rm mid} \le C_{\rm index} < C_{\rm high}$, where $C_{\rm mid}=1$. These objects show moderate anomalies that may have natural dynamical explanations but are nonetheless interesting.
\item \textbf{Tier~3 :} clusters with $0 \le C_{\rm index} < C_{\rm mid}$.
\item \textbf{Tier~4:} clusters with $C_{\rm index} = 0$, these define the control population used to benchmark the behaviour of the metrics and to host the injection experiments described in Section~\ref{sec:injections}.
\end{itemize}
We emphasize that the exact numerical values of the tier thresholds are not critical for the main conclusions; in Section~\ref{sec:results} we show that the identity of the highest--$C$ clusters is robust to modest variations in these choices.

\subsection{Robustness checks and alternative normalizations}
\label{subsec:robustness}

Finally, we test the robustness of the crystallization index against two potential sources of bias: the relative weighting of radial versus kinematic components and the dependence of $C_{\rm index}$ on the total number of stars contributing to each cluster, as well as the choice of radial binning.

First, we explore a family of modified indices of the form
\begin{equation}
C_f \equiv
\left[
(f\,z_{\rm rad,pos})^2 + z_{\rm vel,pos}^2
\right]^{1/2},
\label{eq:C_f_def}
\end{equation}
with $f$ in the range $0.5$--$2$. This effectively reweights the importance of radial structure relative to velocity structure. We find that the rank ordering of clusters by $C_f$ is extremely stable across this range of $f$, and that the subset of highest--$C$ clusters is essentially unchanged; we therefore adopt $f=1$ as our fiducial choice.

Second, because clusters with more stars naturally permit more precise measurements of structure, we explicitly ''de--trend'' the velocity component with sample size. In practice we work with the same $Q_{\rm vel,local}$ values and total star counts $N_\star$ (the number of stars in the working core sample with
$0.5 \le R_{\rm norm} \le 3.0$) that enter the ensemble calibration of
$z_{\rm vel}$ (Section~\ref{subsec:zvel_def}), and fit a simple linear
relation of the form
\begin{equation}
    \log_{10} Q_{\rm vel,local} \;=\; a + b \,\log_{10} N_\star
\end{equation}
across the full cluster sample. We then form residuals
\begin{equation}
    \Delta \log_{10} Q \;\equiv\; \log_{10} Q_{\rm vel,local} - \left(a + b \,\log_{10} N_\star\right)
\end{equation}
and normalize them by their scatter $\sigma_{\rm resid}$ to obtain an $N_\star$--corrected velocity $z$--score,
\begin{equation}
    z_{\rm vel,resid} \;\equiv\; \frac{\Delta \log_{10} Q}{\sigma_{\rm resid}} \, , \qquad
    z_{\rm vel,resid,pos} \;\equiv\; \max\left(0,\,z_{\rm vel,resid}\right) \, .
\end{equation}
The corresponding $N_\star$--corrected crystallization index used in Section~\ref{subsec:C_vs_env} is then
\begin{equation}
    C_{\rm resid} \;\equiv\;
    \left[
        z_{\rm rad,pos}^2 + z_{\rm vel,resid,pos}^2
    \right]^{1/2}.
    \label{eq:Cresid_def}
\end{equation}
In Section~\ref{subsec:C_vs_env} we show that this procedure removes most of the monotonic trend of $Q_{\rm vel,local}$ (and hence $C_{\rm index}$) with $N_\star$, while the clusters identified as having the highest crystallization remain clear outliers. For the purposes of defining tiers and conducting injection tests we therefore use the uncorrected $C_{\rm index}$ as our primary ranking metric, and treat $C_{\rm resid}$ as a diagnostic for environmental correlations. \textcolor{red}{We additionally use the fixed-$N$ downsampling experiment described in Section~\ref{subsec:C_vs_N} as a direct sensitivity check on the impact of sample size, and use $\log_{10}(t_{\rm age}/t_{\rm rh})$ as a simple dynamical-age proxy when comparing clusters of different relaxation state.}

\section{Results}
\label{sec:results}

\subsection{Global distribution of radial and kinematic structure}
\label{subsec:global_C}

\begin{figure*}
\centering
\includegraphics[width=0.6\textwidth]{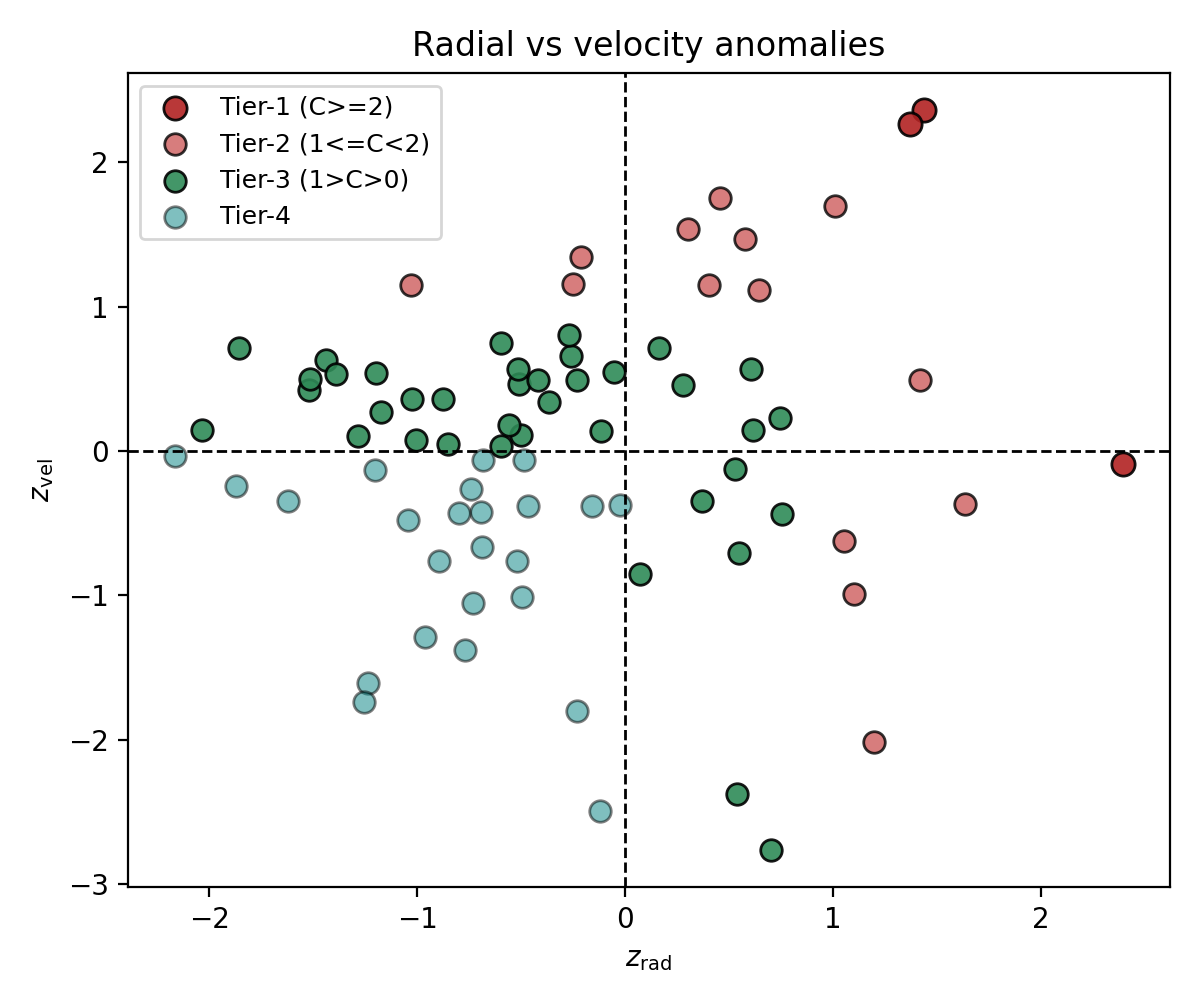}
\caption{
Joint distribution of radial and kinematic structure metrics for the 79 clusters in our sample. Each point shows a cluster in the $(z_{\rm rad},z_{\rm vel})$ plane, where $z_{\rm rad}$ is the standardized radial lumpiness statistic defined in Equation~(\ref{eq:z_rad_def}) and $z_{\rm vel}$ is the standardized local tangential--velocity statistic defined in Equation~(\ref{eq:z_vel_def}). Points are color--coded by crystallization tier based on $C_{\rm index}$ (Section~\ref{subsec:Cindex_def}). \textcolor{red}{Tier~4 controls lie mostly in the lower-left/negative quadrant because $C_{\rm index}$ clips negative $z$--scores to zero; they are therefore smooth relative to both null references rather than necessarily located at $(0,0)$. Tier~1 clusters appear as clear outliers at large positive $z_{\rm vel}$ and/or $z_{\rm rad}$, while one-sided outliers highlight whether the dominant anomaly is kinematic or spatial.}}
\label{fig:zvel_zrad}
\end{figure*}

We first examine the joint distribution of radial and kinematic structure across the full sample of 79 clusters, as quantified by the standardized metrics $z_{\rm rad}$ and $z_{\rm vel}$ defined in Section~\ref{sec:metrics}. Figure~\ref{fig:zvel_zrad} shows $z_{\rm vel}$ versus $z_{\rm rad}$ for all clusters, with points color--coded by the crystallization tier defined from $C_{\rm index}$ (Section~\ref{subsec:Cindex_def}). The majority of clusters lie near the origin, with $|z_{\rm rad}|\lesssim 1$ and $|z_{\rm vel}|\lesssim 1$, indicating levels of radial and local velocity structure consistent with the null expectations calibrated in Sections~\ref{subsec:zrad_def} and \ref{subsec:zvel_def}. 

\begin{figure}
\centering
\includegraphics[width=0.48\textwidth]{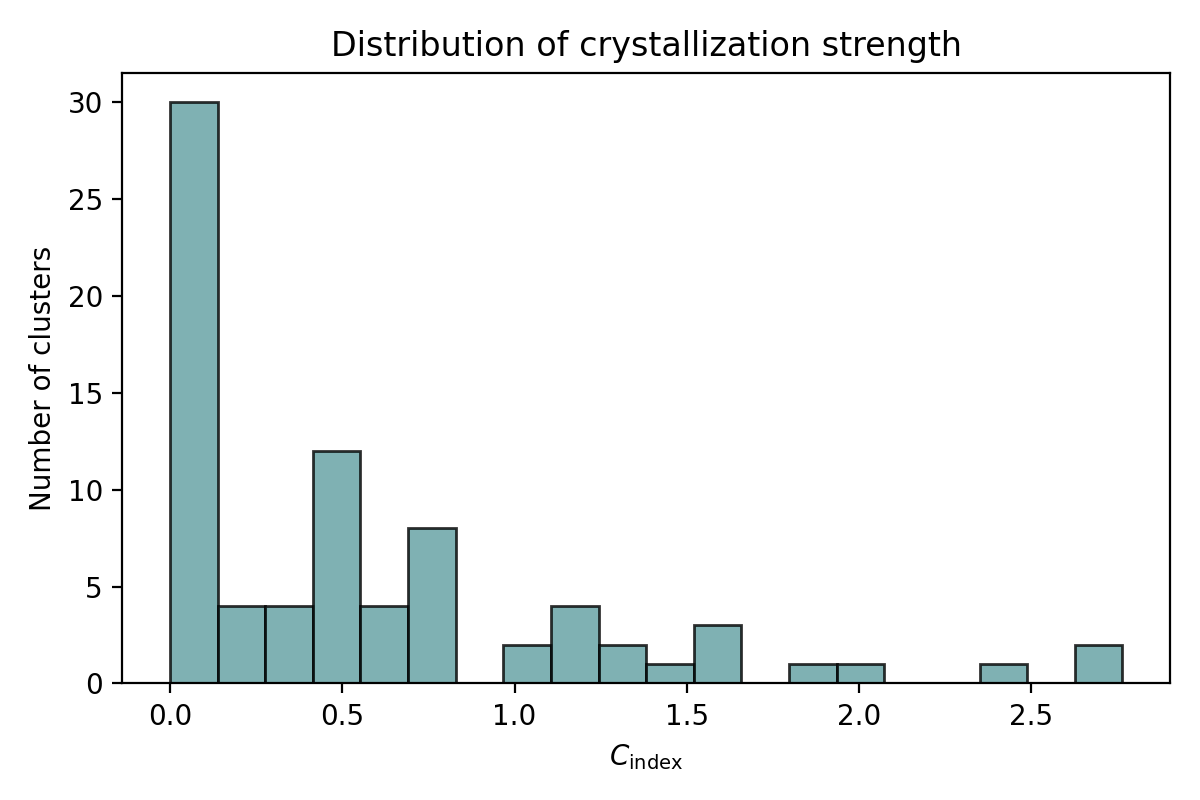}
\caption{
Distribution of the crystallization index $C_{\rm index}$ across the GC sample. The histogram peaks near $C_{\rm index}\approx 0$, indicating that most clusters have radial and local velocity structure consistent with smooth null expectations, and exhibits a long tail towards high $C_{\rm index}$ populated by a small set of outliers. Vertical lines illustrate the tier boundaries adopted in Section~\ref{subsec:Cindex_def}.}
\label{fig:hist_C}
\end{figure}

\textcolor{red}{A small set of objects stand out as clear outliers. These high--$C$ systems populate the upper half of the plane, with $z_{\rm vel}\gtrsim 2$ and, in some cases, moderately elevated $z_{\rm rad}$. Well--known dynamically complex clusters such as NGC~5139 (Omega Centauri) and NGC~104 (47 Tucanae) fall into this category, as does the system BH~140. The high velocity components for NGC~104 and NGC~5139 are qualitatively consistent with independent proper-motion studies that find non-Gaussian moment structure, including skewness and population-dependent kinematics \citep{Heyl2017_47Tuc,Ziliotto2025_47TucKinematics,Ziliotto2026OmegaCen}. BH~140, by contrast, is primarily radial in our metric ($z_{\rm rad}>0$, $z_{\rm vel}<0$), illustrating that a high $C_{\rm index}$ can arise from spatial inhomogeneity without an anomalous tangential-speed distribution. Conversely, clusters with large $z_{\rm vel}$ but modest $z_{\rm rad}$ are candidates for velocity-distribution complexity without a sharp radial overdensity. In our classification scheme these objects are assigned to Tier~1. The one--dimensional distribution of the combined crystallization index $C_{\rm index}$ is shown in Figure~\ref{fig:hist_C}. The histogram peaks sharply near $C_{\rm index}\approx 0$, with a long tail extending to $C_{\rm index}\gtrsim 2$. Most clusters occupy the control regime (Tier~4), with $C_{\rm index}\equiv 0$, while Tier~3 spans $0< C_{\rm index}<1$ and Tier~2 spans $1< C_{\rm index}<2$. The Tier~1 clusters inhabit the extreme tail. This behaviour is consistent with a picture in which a largely smooth GC population is punctuated by a handful of systems with strongly enhanced phase--space structure. The complete table listing all the crystallization information is shown in Tables~\ref{tab:tiers1} and \ref{tab:tiers2}.}

To quantify the statistical significance of this high--$C$ tail, we compare the observed number of clusters above the Tier~1 threshold to the expectation from an analytic null. Approximating $z_{\rm rad}$ and $z_{\rm vel}$ as independent standard normal variables with positive parts contributing to $C_{\rm index}$ (Equation~\ref{eq:Cindex_def}), we derive an effective single--cluster probability $p_0 \simeq 0.0566$ of having $C_{\rm index}\ge C_{\rm high}=2$. For 79 clusters this implies an expected count of $N_{\rm high} = 4.47\pm 2.05$ with $C_{\rm index}\ge 2$. In the data we observe three such clusters. A Monte Carlo experiment drawing $10^5$ ensembles of 79 clusters from the analytic null yields a fraction $\simeq 8.3\times 10^{-1}$ of realizations with $N_{\rm high}\ge 3$, indicating that the observed high--$C$ tail is fully consistent with the smooth baseline and does not constitute a statistically significant excess. In what follows we therefore treat the Tier~1 systems as the high end of the crystallization distribution, without claiming a global departure from the analytic null model.

\subsection{Crystallization versus sample size}
\label{subsec:C_vs_N}

\begin{figure}
\centering
\includegraphics[width=0.48\textwidth]{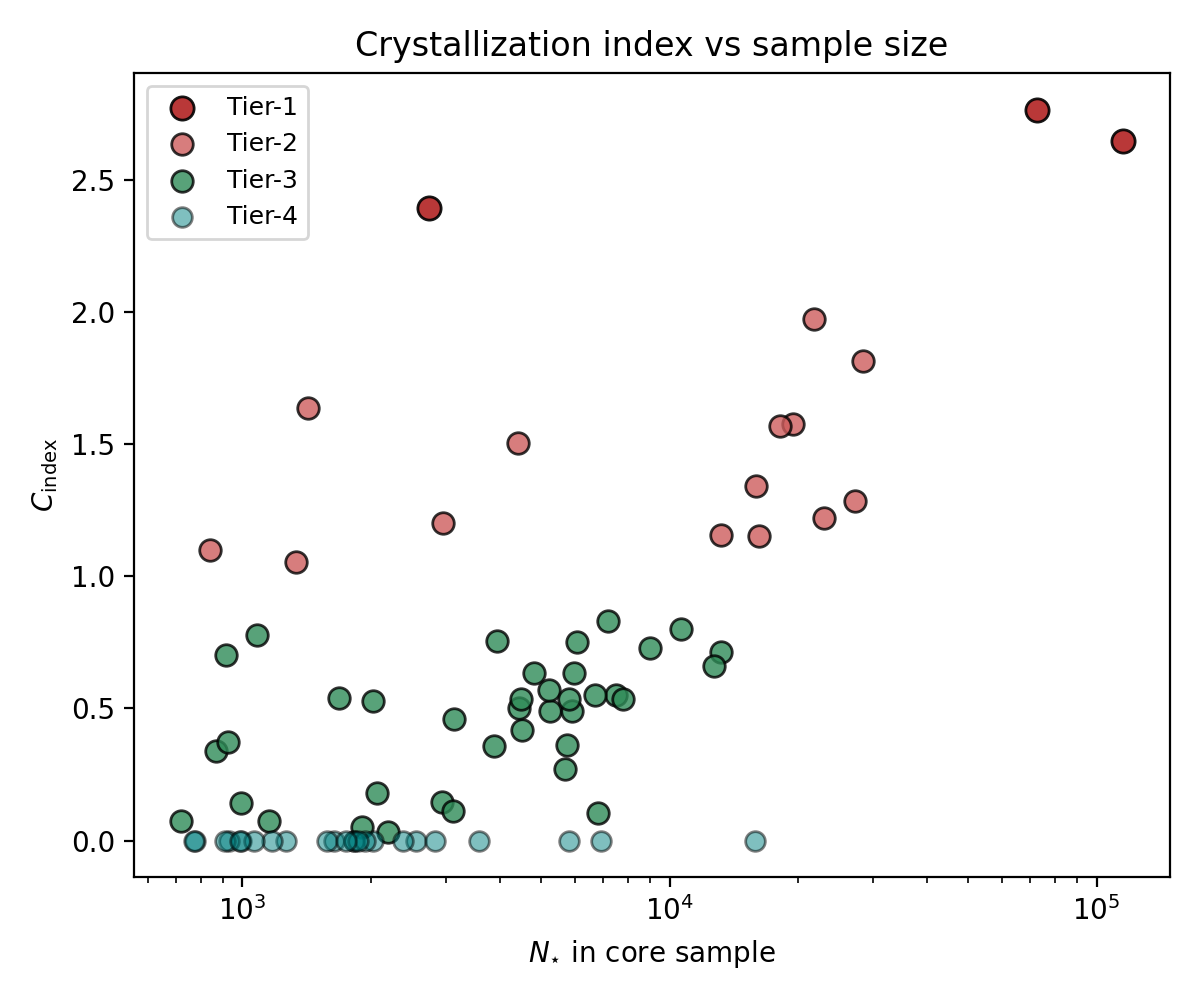}
\caption{
Crystallization index $C_{\rm index}$ as a function of the number of analyzed core stars $N_{\rm core}$ for all clusters. \textcolor{red}{The positive trend shows that sample size is an important sensitivity driver. We therefore use the residual index $C_{\rm resid}$ and fixed-$N$ subsampling as complementary diagnostics rather than interpreting raw $C_{\rm index}$ as a sample-size-free intrinsic quantity.}}
\label{fig:C_vs_N}
\end{figure}

\textcolor{red}{Because the precision with which small--scale structure can be measured improves with the number of stars, we examine how $C_{\rm index}$ and its kinematic component behave as a function of the analyzed core star count $N_{\rm core}$. Here $N_{\rm core}$ is the final number of stars passing all cuts and entering the metrics. Figure~\ref{fig:C_vs_N} shows $C_{\rm index}$ versus $N_{\rm core}$ for all clusters. There is a clear positive trend: Spearman's rank coefficient between $C_{\rm index}$ and $\log_{10}N_{\rm core}$ is $\rho=0.57$ ($p=3.1\times10^{-8}$). We therefore interpret uncorrected $C_{\rm index}$ as a ranking statistic whose sensitivity is partly sample-size dependent, not as a purely intrinsic structural parameter. The $N_\star$-residual index $C_{\rm resid}$ defined in Equation~\ref{eq:Cresid_def} removes this trend to first order ($\rho=-0.11$, $p=0.32$ for $C_{\rm resid}$ versus $\log_{10}N_{\rm core}$).}

\begin{figure}
\centering
\includegraphics[width=0.48\textwidth]{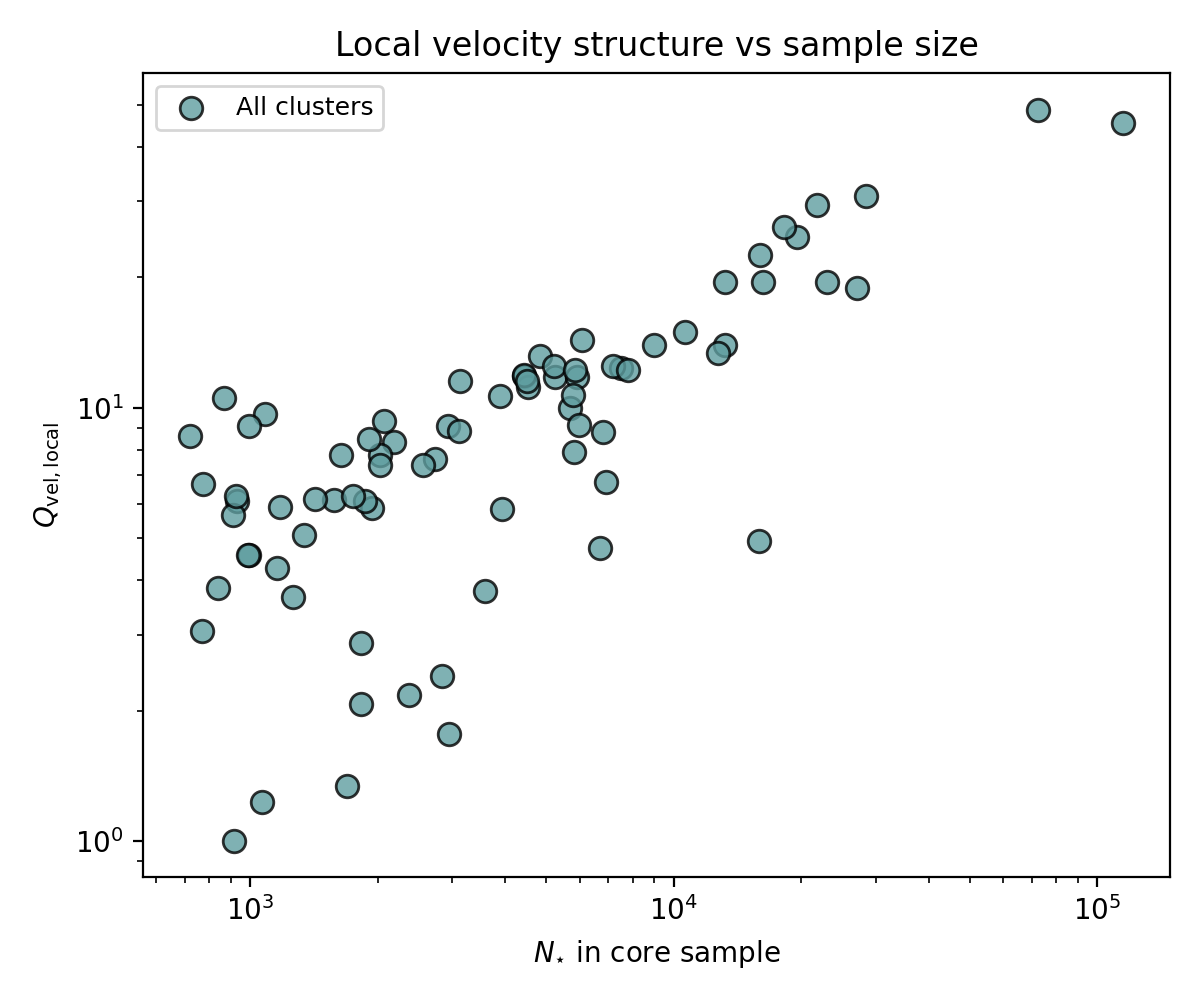}
\caption{
Local tangential--velocity structure metric $Q_{\rm vel,local}$ versus $N_{\rm core}$. \textcolor{red}{The monotonic increase of $Q_{\rm vel,local}$ with sample size reflects improved detectability of small departures from the Rayleigh reference. This trend motivates the $N_\star$-corrected residual velocity score used to construct $C_{\rm resid}$ (Equation~\ref{eq:Cresid_def}).}}
\label{fig:Qvel_vs_N}
\end{figure}

\begin{figure}
\centering
\includegraphics[width=0.48\textwidth]{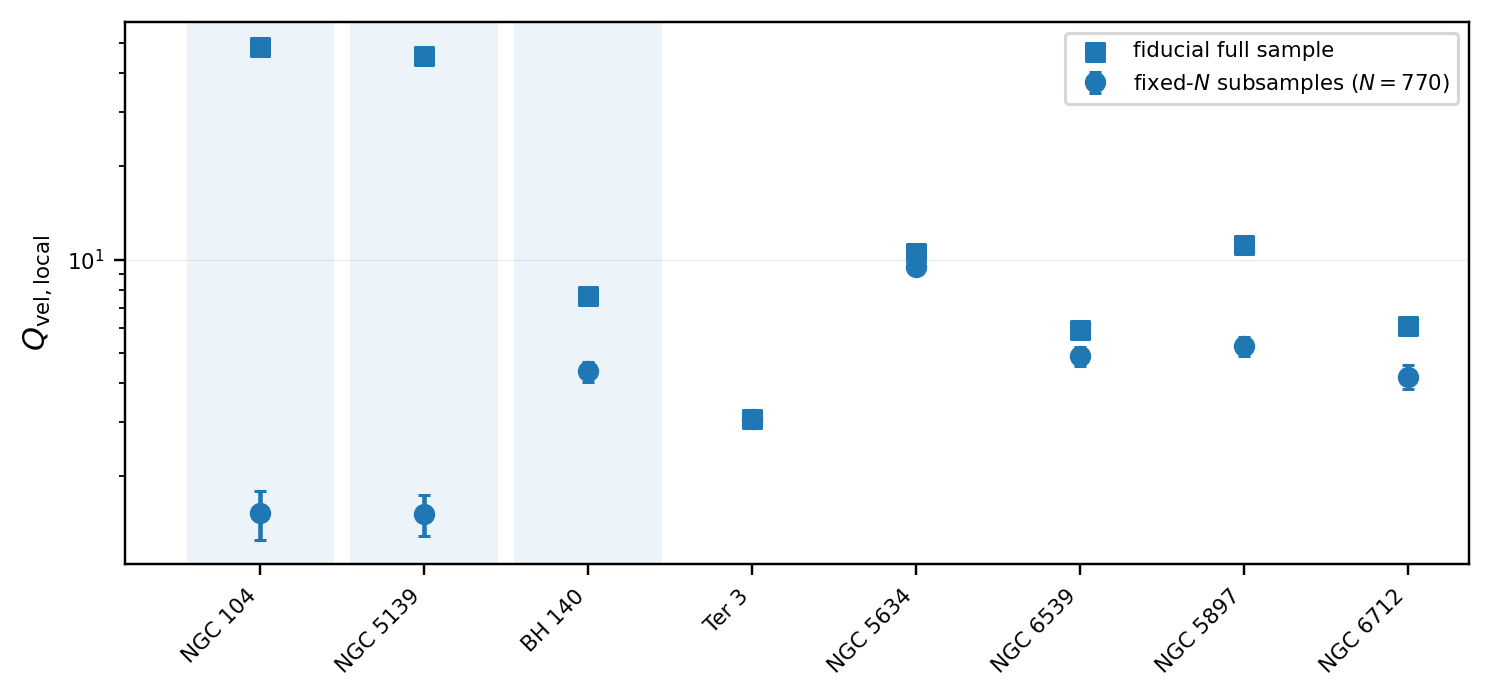}
\caption{\textcolor{red}{Fixed-$N$ downsampling diagnostic for the Tier~1 clusters and representative comparison clusters. Squares show the fiducial full-sample $Q_{\rm vel,local}$ values, circles with error bars show the mean and standard deviation from random subsamples of $N=770$ stars, the smallest analyzed sample size. The large reduction for NGC~104 and NGC~5139 demonstrates the role of sample size in velocity-structure detectability; BH~140 remains relatively elevated because its high $C_{\rm index}$ is dominated by the radial component.}}
\label{fig:fixedN_downsample}
\end{figure}

\textcolor{red}{To disentangle kinematic structure from sample-size effects more directly, Figure~\ref{fig:Qvel_vs_N} shows the local tangential--velocity metric $Q_{\rm vel,local}$ as a function of $N_{\rm core}$. As expected from its weighted-RMS construction (Equation~\ref{eq:Qvellocal_def}), $Q_{\rm vel,local}$ increases with sample size: when more stars are available, the observed velocity distribution samples more of the Rayleigh tail and small deviations are measured with smaller Poisson uncertainty. We therefore report both the uncorrected $C_{\rm index}$ ranking and the residual diagnostic $C_{\rm resid}$. As a complementary test, Figure~\ref{fig:fixedN_downsample} shows fixed-$N$ downsampling to $N=770$ stars for the Tier~1 objects and several comparison clusters. The rich clusters NGC~104 and NGC~5139 lose most of their high $Q_{\rm vel,local}$ amplitude after such severe downsampling, demonstrating that sample size controls detectability. BH~140, whose anomaly is primarily radial, remains comparatively elevated. This test supports a conservative interpretation: high $C_{\rm index}$ identifies objects for follow-up, but the metric should always be read together with $N_{\rm core}$ and $C_{\rm resid}$.}

\subsection{Dependence on global cluster properties}
\label{subsec:C_vs_env}

\begin{figure*}
\centering
\includegraphics[width=0.95\textwidth]{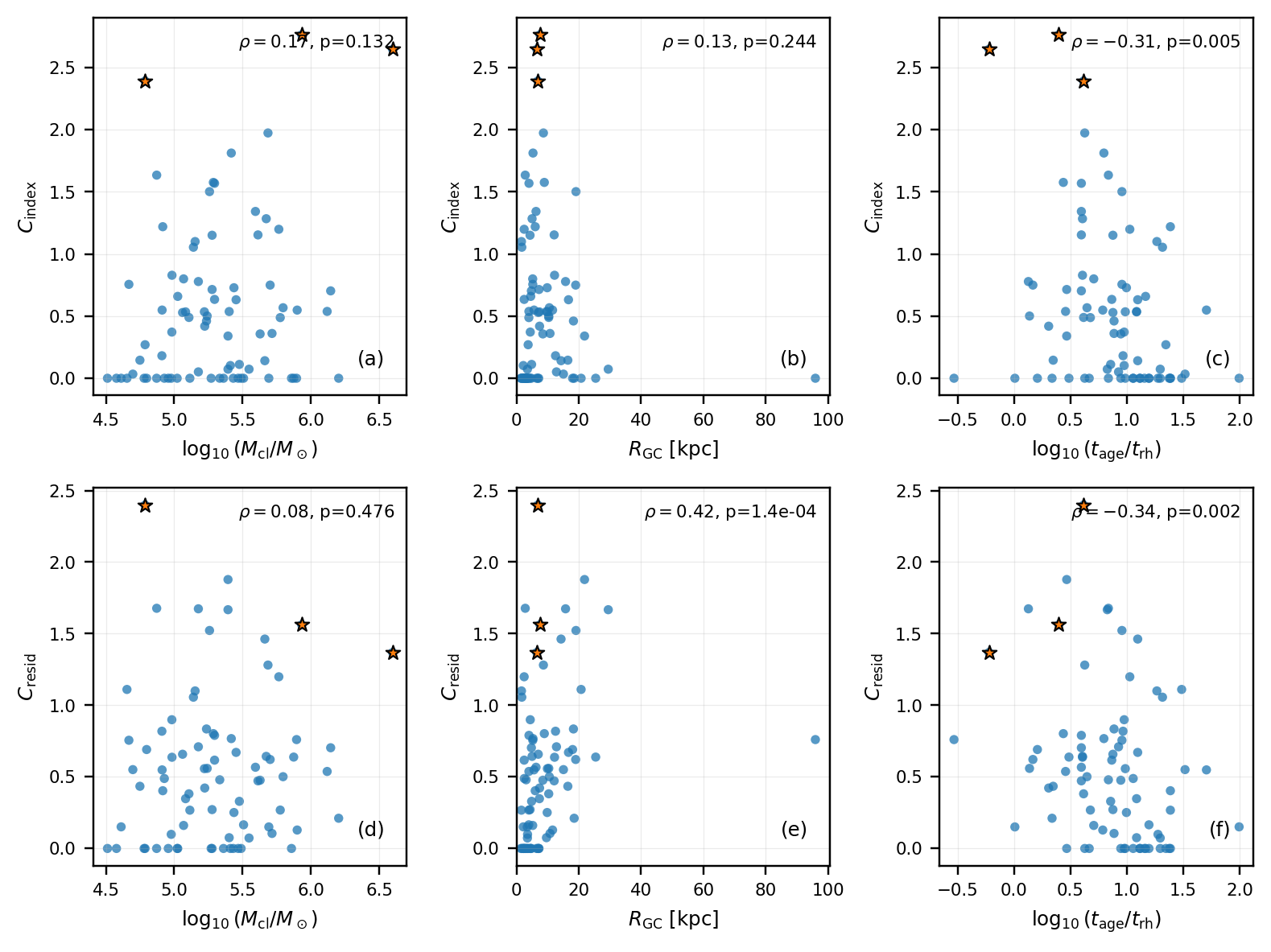}
\caption{\textcolor{red}{Correlations of crystallization metrics with cluster mass, Galactocentric distance, and a dynamical-age proxy. Top row: $C_{\rm index}$ versus $\log(M_{\rm cl}/M_\odot)$, $R_{\rm GC}$, and $\log_{10}(t_{\rm age}/t_{\rm rh})$. Bottom row: the $N_\star$-corrected residual index $C_{\rm resid}$ versus the same quantities. Spearman rank coefficients and two-sided $p$-values are printed in each panel. Star symbols mark Tier~1 clusters.}}
\label{fig:C_mass_Rgc_grid}
\end{figure*}

\textcolor{red}{We next investigate how the crystallization index and its $N$--corrected version relate to basic global properties of the clusters. Figure~\ref{fig:C_mass_Rgc_grid} summarizes the results for cluster mass, Galactocentric distance, and a simple dynamical-age proxy. We define the latter as $\log_{10}(t_{\rm age}/t_{\rm rh})$, using a representative old-GC age of $t_{\rm age}=12.5$~Gyr and the half-mass relaxation time $t_{\rm rh}$ from the \citet{BaumgardtVasiliev2021} compilation. This proxy is not intended to replace a full dynamical classification, but it follows the physical logic of separating dynamically young, intermediate, and old systems by the number of elapsed relaxation times.}

\textcolor{red}{For the uncorrected index, the correlation with mass is weak in rank statistics: $C_{\rm index}$ versus $\log(M_{\rm cl}/M_\odot)$ has Spearman $\rho=0.17$ and $p=0.13$. The correlation with present-day Galactocentric distance is also weak ($\rho=0.13$, $p=0.24$). The strongest trend among the plotted quantities is with the relaxation-time proxy, where $C_{\rm index}$ decreases toward dynamically older systems ($\rho=-0.31$, $p=4.8\times10^{-3}$), consistent with the expectation that repeated relaxation erases small-scale phase-space structure. When we use $C_{\rm resid}$, the mass trend is further reduced ($\rho=0.08$, $p=0.48$), while the dynamical-age trend remains modest ($\rho=-0.34$, $p=2.2\times10^{-3}$). The residual index also shows a moderate positive rank correlation with $R_{\rm GC}$ ($\rho=0.42$, $p=1.4\times10^{-4}$), indicating that the residualized metric may retain environmental information after the sample-size trend is removed. In all cases, however, the scatter is large: no single global parameter uniquely predicts the high-$C$ clusters.}

\textcolor{red}{Dividing the sample by the same relaxation-count proxy gives median $C_{\rm index}=0.50$ for dynamically young systems ($t_{\rm age}/t_{\rm rh}<3$), 0.53 for intermediate systems ($3\le t_{\rm age}/t_{\rm rh}<10$), and 0.00 for dynamically old systems ($t_{\rm age}/t_{\rm rh}\ge10$). This supports the physical interpretation that crystallization is easier to preserve or detect in less relaxed systems, while also reinforcing that the index is not merely a proxy for cluster mass. We do not show a metallicity correlation because metallicities are not part of the homogeneous data table used in the present analysis.}

\subsection{Tier populations and control sample}
\label{subsec:tier_pop}
\textcolor{red}{The distributions discussed above naturally partition the cluster population into the four tiers defined in Section~\ref{subsec:Cindex_def}. Tier~1 consists of three clusters in the high--$C_{\rm index}$ tail ($C_{\rm index}\gtrsim 2$). These clusters are prime targets for more detailed dynamical modelling because they are metric outliers; the tier label should not be read as evidence for non-standard physics or engineering. Known mechanisms such as multiple populations, rotation, anisotropy, tides, accretion history, or catalogue systematics remain the default explanations to test first.}

Tier~2 and 3 comprises 52 clusters with intermediate and low crystallization indices ($0< C_{\rm index}<2$). Many of these show modest enhancements in either $z_{\rm rad}$ or $z_{\rm vel}$ individually, but do not stand out as strongly as the Tier~1 systems in the joint $(z_{\rm rad},z_{\rm vel})$ plane. While their structure is likely explicable in terms of known dynamical processes (e.g.\ internal rotation, recent tidal perturbations, or residual substructure from formation), they remain useful comparison objects.

Tier~4 collects 24 clusters with $C_{\rm index}= 0$, including a subset of 10 (control group). The latter systems have both $z_{\rm rad}\le 0$ and $z_{\rm vel}\le 0$ and behave, within uncertainties, like smooth null realizations of their radial and local velocity distributions. These make up our control pool for the injection tests described in Section~\ref{sec:injections}. By adding artificial ultra--cold components of varying strength to Tier~4 clusters and re--evaluating $C_{\rm index}$, we are able to quantify the sensitivity of our method and to translate the non--detection of high crystallization in these controls into upper limits on any ultra--cold kinematic substructure.

\section{Injection Tests and Detection Limits}
\label{sec:injections}

The crystallization index $C_{\rm index}$ provides a relative ranking of clusters by their phase--space structure, but by itself it does not directly answer the question: \emph{how sensitive is our method to a genuinely cold, coherent component, if such a component were present?} In this section we address that question by performing controlled injection experiments in dynamically smooth Tier~4 clusters, and by translating the absence of high crystallization in these controls into quantitative upper limits on ultra--cold kinematic substructures.

\subsection{Signal model and experimental setup}
\label{subsec:signal_model}

We model a hypothetical crystallized component as a population of stars confined to a limited radial range and occupying an extremely narrow region of tangential velocity space. Motivated by the construction of $Q_{\rm vel,local}$ in Section~\ref{subsec:Qvellocal_def}, we work in units of the shell--level dispersion $\sigma_s$ and define
\begin{equation}
x \equiv \frac{v_{\rm tan}}{\sigma_s} \,,
\end{equation}
where $v_{\rm tan}$ is the tangential speed in a given radial shell and $\sigma_s$ is the robust IQR--based dispersion of the original shell velocities (before any injection). Under the null hypothesis of an isotropic, Gaussian velocity field, the distribution of $x$ is approximately Rayleigh with unit scale. A cold, crystallized component is therefore characterized by a strongly peaked excess at $x \ll 1$.

In our injection experiments we consider an idealized signal of the form
\begin{equation}
p_{\rm crystal}(x) \propto
\exp\left[-\frac{(x - x_0)^2}{2\sigma_x^2}\right] \,,
\end{equation}
with $x_0 \approx 0.1$ and $\sigma_x \approx 0.01$, i.e.\ a very narrow spike in normalized tangential speed close to zero. We confine this component to the intermediate radial shell,
\begin{equation}
{\rm shell~2}: \quad 1.0 \le R_{\rm norm} < 2.0 \,,
\end{equation}
\textcolor{red}{because this shell is the best-populated radial region in the adopted samples and avoids the most severe central crowding and outer-field contamination. It is therefore the most favorable single-shell location for a first sensitivity experiment, not a claim that real substructures must preferentially form there.} Within shell~2 we select a fraction $f$ of the stars and replace their original $v_{\rm tan}$ values by draws from $p_{\rm crystal}(x)$, converted back to physical units using the shell's original $\sigma_s$. The remaining stars in the shell, and all stars in shells~1 and 3, are left unchanged.

For each choice of $f$ we recompute $Q_{\rm vel,local}$, $\log_{10}Q_{\rm vel,local}$, $z_{\rm vel}$, and the resulting crystallization index. We perform this procedure for a grid of signal strengths $f$ between 0 and 1. In practice we adopt a small discrete set of values (e.g.\ $f=0,\,0.05,\,0.10,\,0.20,\,0.40,\,0.60,\,0.80,\,1.0$), since the dependence of the metrics on $f$ is smooth on these scales.

\subsection{Global and control--ensemble normalization}
\label{subsec:global_vs_control}

The same injected signal can appear more or less anomalous depending on the reference ensemble used to define $z_{\rm vel}$ and $C_{\rm index}$. To capture this dependence we construct two parallel normalizations:
\begin{itemize}[nosep,leftmargin=*]
\item \textbf{Global ensemble:}  
  We use the full set of 79 clusters to define the mean and dispersion of $\log_{10}Q_{\rm vel,local}$,
  \begin{equation}
  \mu_{\log Q,{\rm all}} =
  \left\langle \log_{10} Q_{\rm vel,local} \right\rangle_{\rm all}, \quad
  \sigma_{\log Q,{\rm all}} =
  {\rm std}_{\rm all}\left(\log_{10}Q_{\rm vel,local}\right),
  \end{equation}
  and compute
  \begin{equation}
  z_{\rm vel,all} =
  \frac{\log_{10}Q_{\rm vel,local} - \mu_{\log Q,{\rm all}}}
       {\sigma_{\log Q,{\rm all}}}
  \end{equation}
  for the injected cluster. The corresponding crystallization index,
  \begin{equation}
  C_{\rm all} =
  \left[
    z_{\rm rad,pos}^2 + z_{\rm vel,all,pos}^2
  \right]^{1/2},
  \end{equation}
  measures how strongly the injected cluster stands out \emph{relative to the full GC population}.
\item \textbf{Control ensemble:}  
  We restrict the reference set to the Tier~4 control clusters (Section~\ref{subsec:tier_pop}), which by construction have $C_{\rm index}\lesssim 1$ and include a subset with $C_{\rm index}\approx 0$. Using only these controls we define
  \begin{equation}
  \mu_{\log Q,{\rm ctrl}} =
  \left\langle \log_{10} Q_{\rm vel,local} \right\rangle_{\rm ctrl}, \quad
  \sigma_{\log Q,{\rm ctrl}} =
  {\rm std}_{\rm ctrl}\left(\log_{10}Q_{\rm vel,local}\right),
  \end{equation}
  and compute
  \begin{equation}
  z_{\rm vel,ctrl} =
  \frac{\log_{10}Q_{\rm vel,local} - \mu_{\log Q,{\rm ctrl}}}
       {\sigma_{\log Q,{\rm ctrl}}}.
  \end{equation}
  The corresponding
  \begin{equation}
  C_{\rm ctrl} =
  \left[
    z_{\rm rad,pos}^2 + z_{\rm vel,ctrl,pos}^2
  \right]^{1/2}
  \end{equation}
  quantifies how anomalous the injected cluster would appear \emph{within a population that is otherwise smooth}.
\end{itemize}
In both cases we hold the radial structure metric $z_{\rm rad}$ fixed, since the injection modifies only the tangential velocities in shell~2. For Tier~4 controls, $z_{\rm rad,pos}$ is typically zero or small.

\subsection{Choice of control clusters}
\label{subsec:control_choice}

We perform injection tests in six representative Tier~4 clusters selected from the control pool with $C_{\rm index}\approx 0$ (or slightly negative), chosen to span more than an order of magnitude in core sample size and to cover a range of structural properties:
\begin{itemize}[nosep,leftmargin=*]
\item Terzan 3 (Ter~3): a very low--$N$ control with $N_{\rm core}=770$ stars in $0.5\le R_{\rm norm}\le 3.0$. This cluster has strongly negative baseline $z_{\rm vel}$ and $z_{\rm rad}$, and acts as an extreme example of a very sparsely sampled, dynamically smooth core.
\item IC~1276: a low-- to intermediate--$N$ control with $N_{\rm core}=1264$ in the same radial range. Its baseline crystallization index is consistent with zero, but shell~2 still contains enough stars to allow meaningful injections.
\item NGC~6093 (M80): an intermediate--$N$ control with $N_{\rm core}=2552$. This compact bulge cluster provides an example where the core is reasonably well populated but still far from the richest systems.
\item NGC~6838 (M71): another intermediate--$N$ control with $N_{\rm core}=3589$, representative of relatively low--mass, nearby clusters with good \emph{Gaia} sampling.
\item NGC~4833: a high--$N$ halo cluster with $N_{\rm core}=6906$, illustrating the behaviour of our metric in a well--sampled control where the core contains several thousand stars.
\item NGC~6121 (M4): the richest of our injection hosts, with $N_{\rm core}=15\,886$ stars in $0.5\le R_{\rm norm}\le 3.0$. This nearby, very well--sampled cluster provides a best--case scenario for detecting weak ultra--cold components at fixed fractional strength.
\end{itemize}
Taken together, these six clusters bracket the range of star counts present in the Tier~4 control sample and thus provide a realistic view of how sensitivity to crystallized components scales with $N_{\rm core}$ in our \emph{Gaia}--based core catalogues.

\subsection{\texorpdfstring{\textcolor{red}{Response of the crystallization index to injected crystals}}{Response of the crystallization index to injected crystals}}
\label{subsec:C_vs_f}

\begin{figure*}
\centering
\begin{tabular}{ccc}

\includegraphics[width=0.31\textwidth]{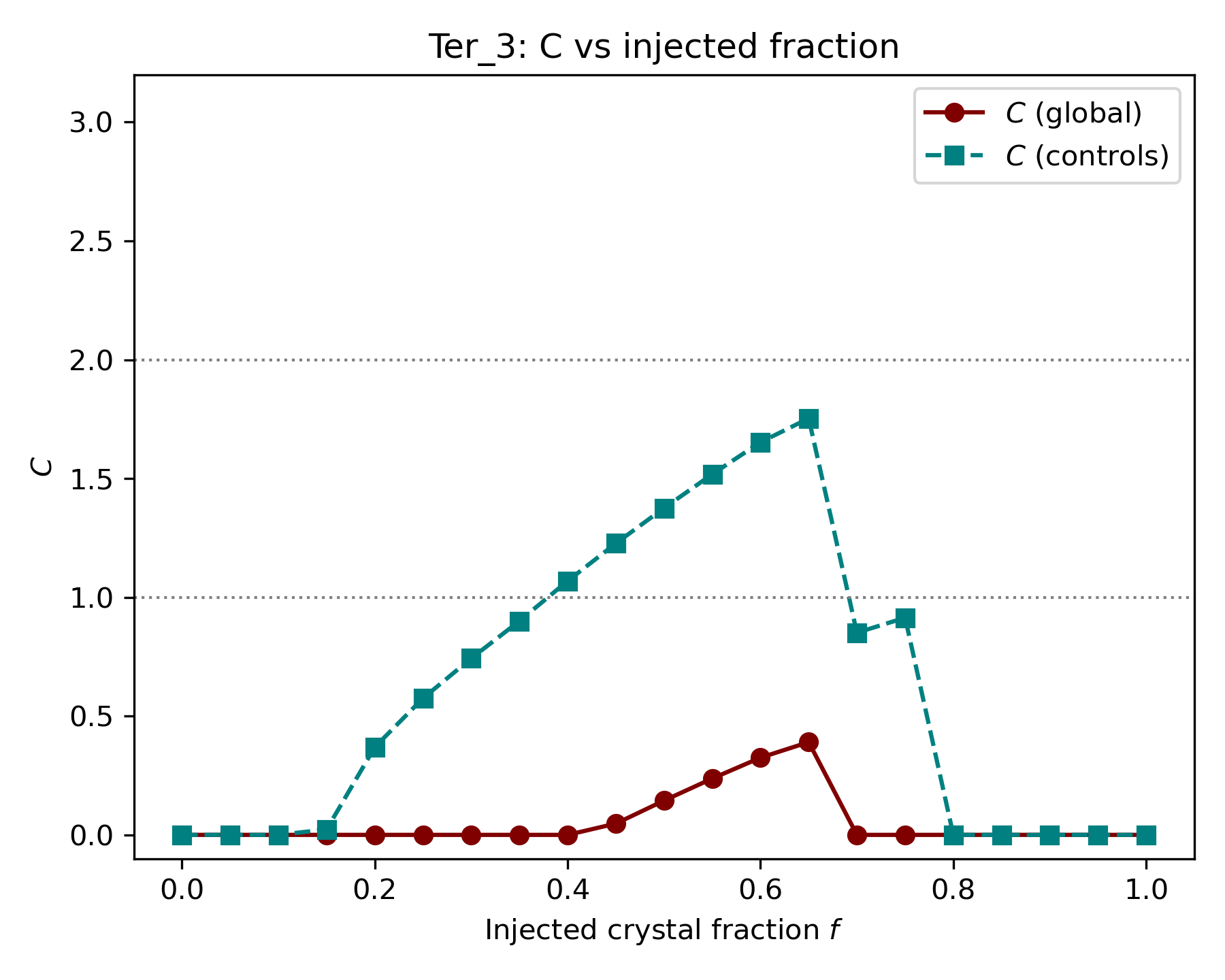}&
\includegraphics[width=0.31\textwidth]{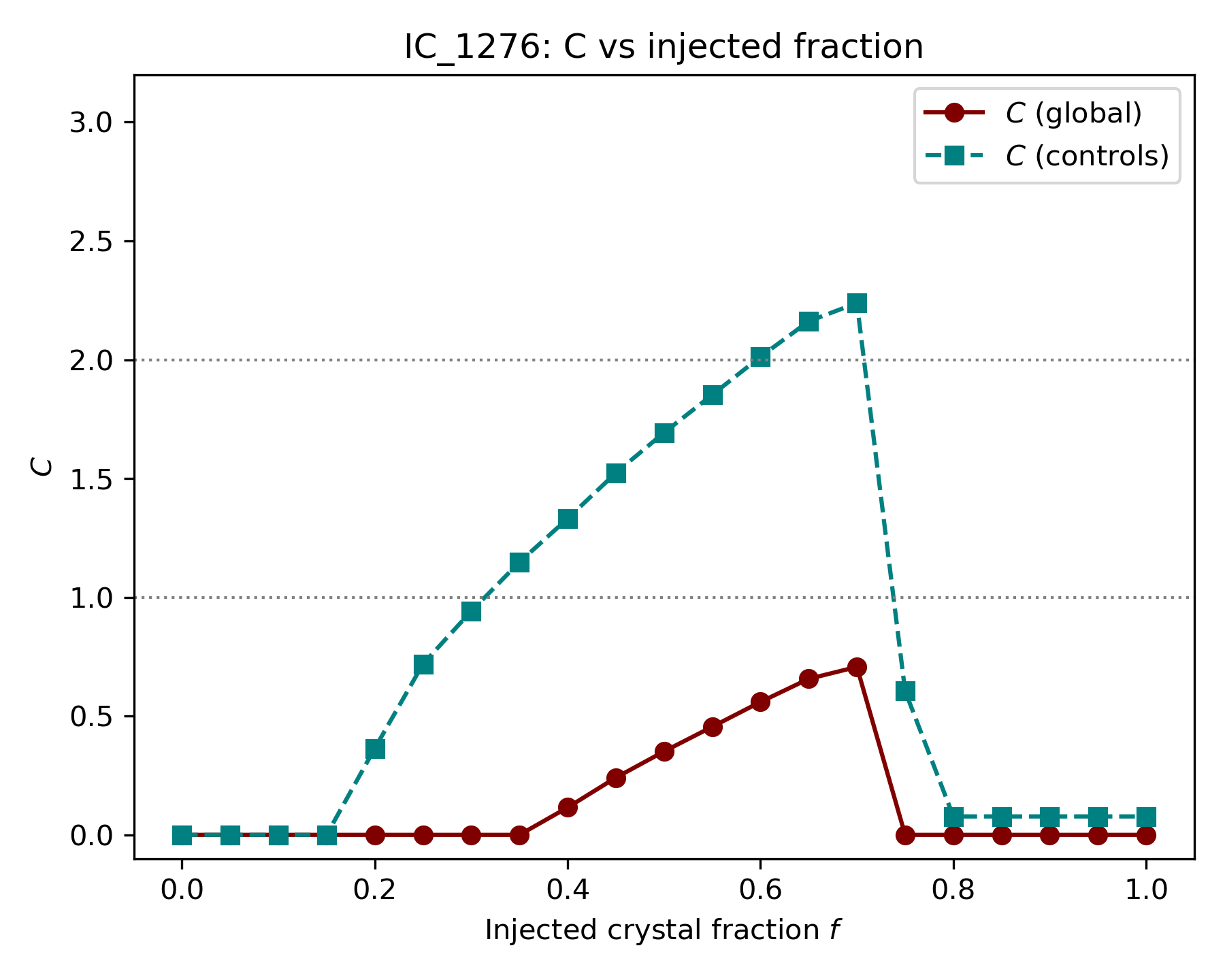}&
\includegraphics[width=0.31\textwidth]{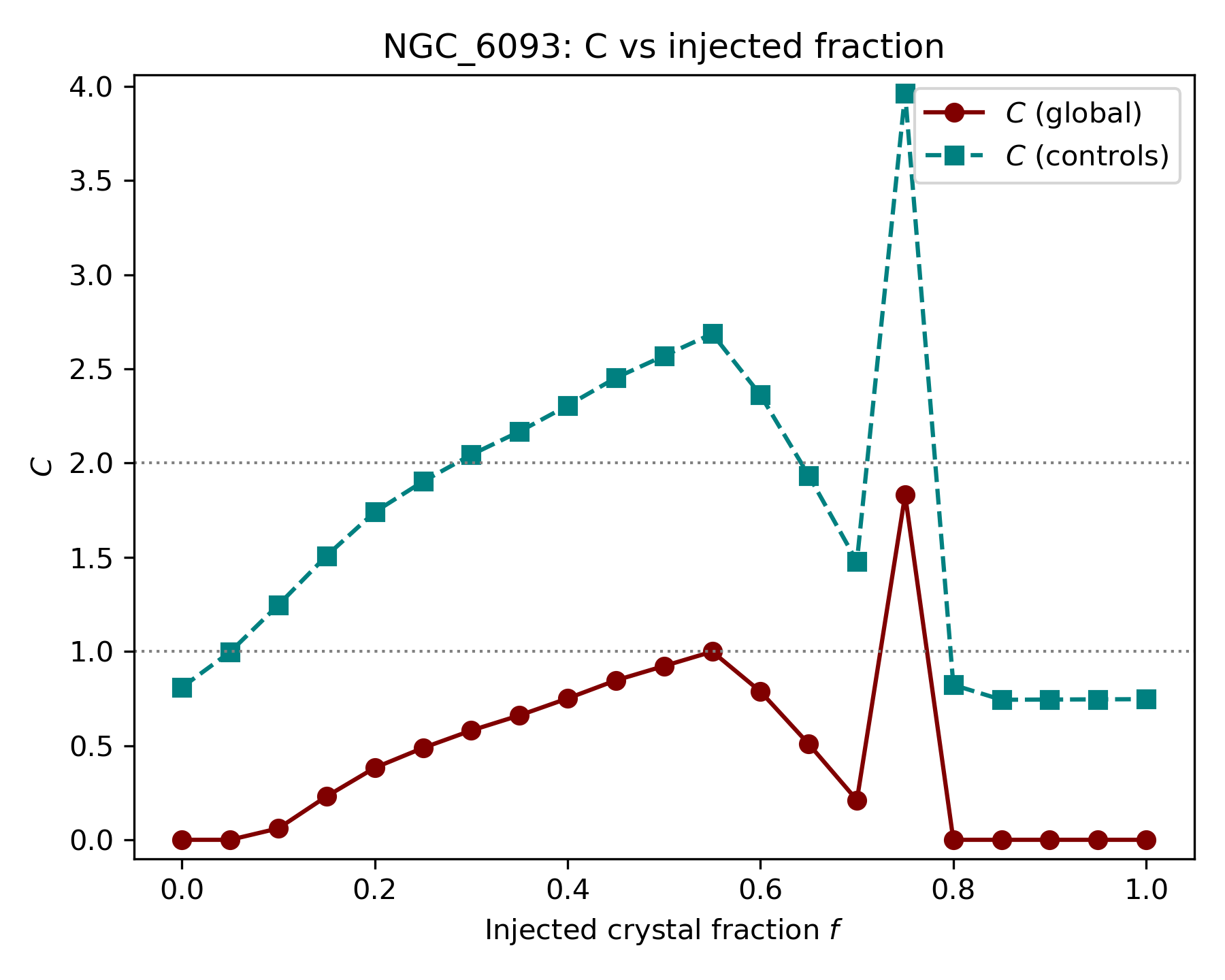}  \\
\includegraphics[width=0.31\textwidth]{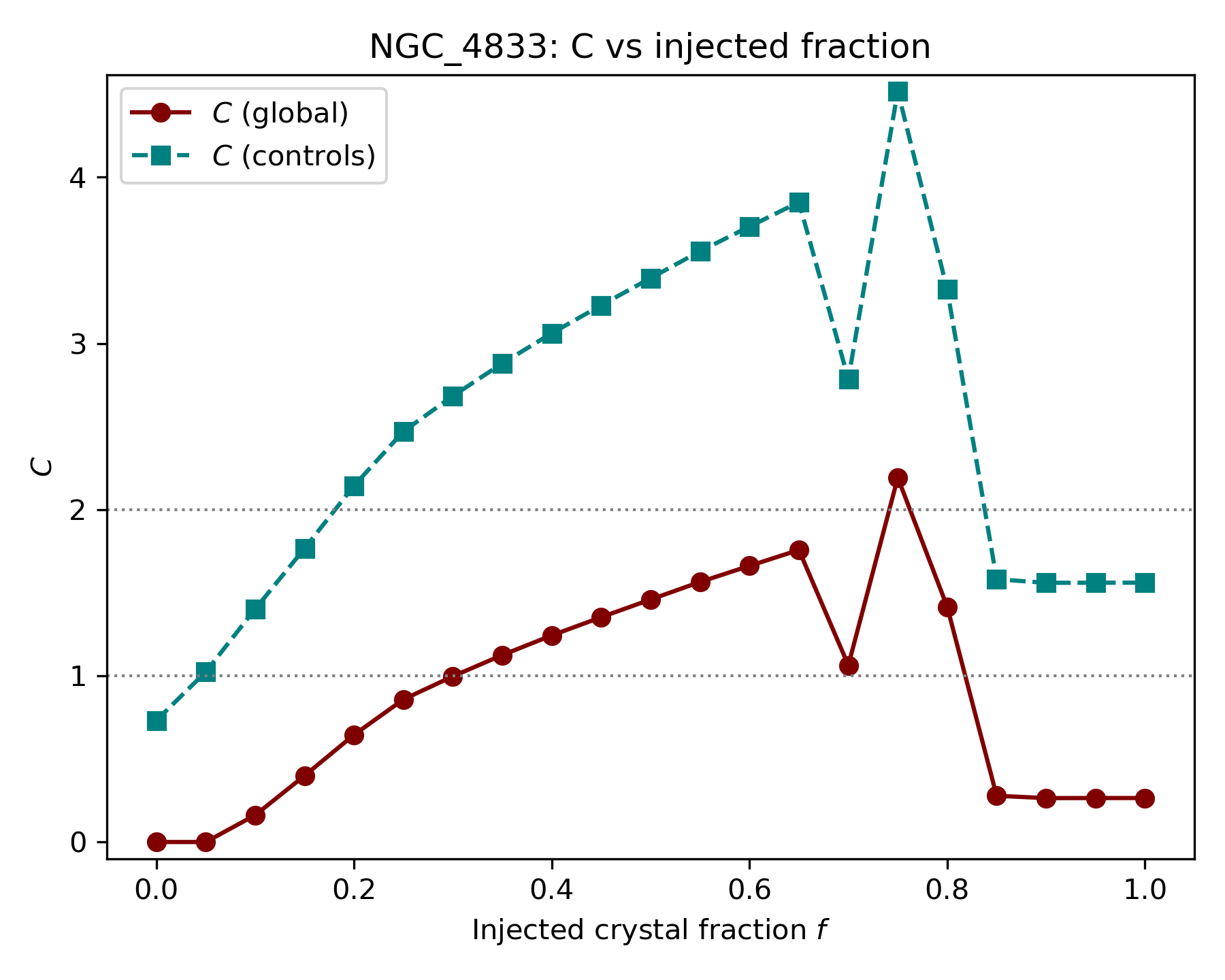} &
\includegraphics[width=0.31\textwidth]{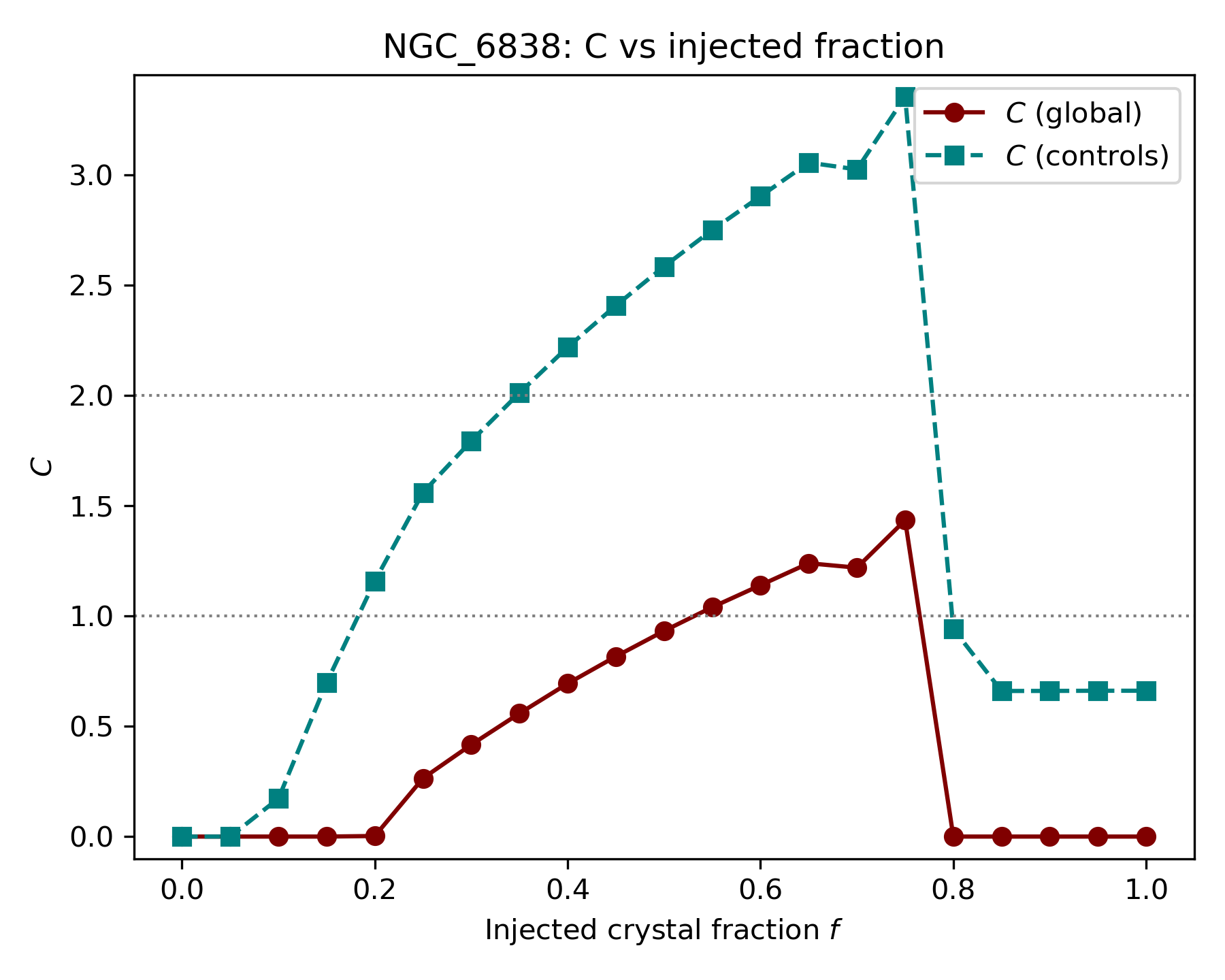} &
\includegraphics[width=0.31\textwidth]{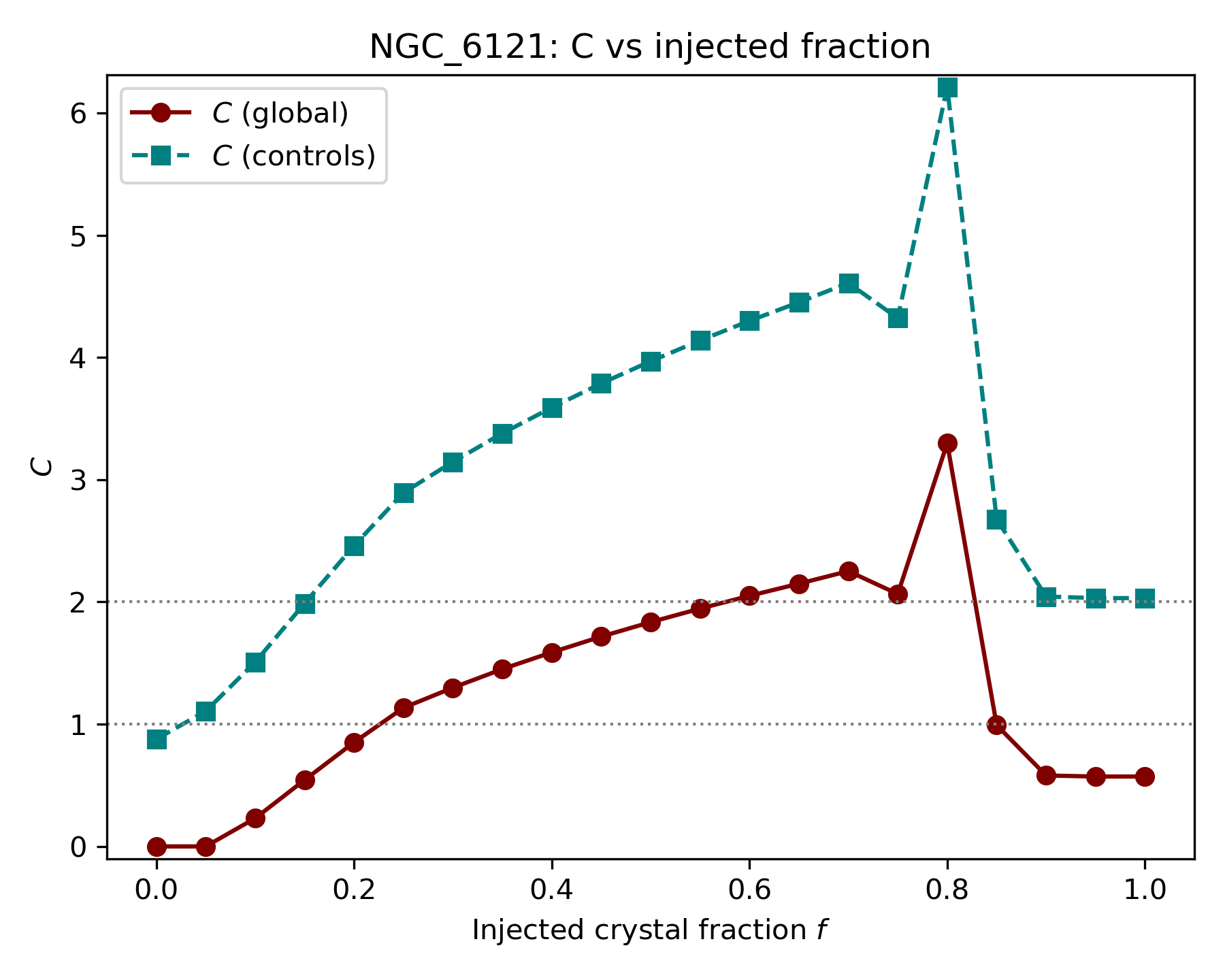} \\

\end{tabular}
\caption{
Response of the crystallization index to injected ultra--cold components in six representative clusters. Each panel shows the median crystallization index $C$ as a function of the injected fraction $f$ of shell--2 stars ($1 \le R_{\rm norm} < 2$) reassigned to the ultra--cold component. Solid circles trace $C_{\rm all}$, where $z_{\rm vel}$ is calibrated against the full 79--cluster ensemble, while dashed squares show $C_{\rm ctrl}$, where $z_{\rm vel}$ is calibrated only against the low--$C$ control clusters. Horizontal dotted lines mark $C=2$ and $C=1$, with $C=2$ adopted as the Tier~1 boundary. \textcolor{red}{The turnover near large $f$ reflects the self-normalization by the current shell dispersion, not disappearance of an injected cold component.}}
\label{fig:C_vs_f_six}
\end{figure*}

Figure~\ref{fig:C_vs_f_six} shows, for each of the six host clusters, how the crystallization index responds to an injected ultra--cold component as a function of the injected fraction $f$ of shell--2 stars.  In every panel the solid circles trace the ''global'' index $C_{\rm all}$, where $z_{\rm vel}$ is calibrated using the full 79--cluster ensemble, while the dashed squares show the ''control'' index $C_{\rm ctrl}$, where $z_{\rm vel}$ is calibrated only against the smooth, Tier~4 population.

At small and intermediate $f$ the behaviour is qualitatively similar in all hosts.  As more stars in shell~2 are reassigned to the ultra--cold component, the local tangential--velocity histogram becomes increasingly bimodal, with a narrow peak near $v_{\rm tan}\simeq 0$ superimposed on a broader ''thermal'' background.  This drives $Q_{\rm vel,local}$ above its baseline value and causes both $C_{\rm all}$ and $C_{\rm ctrl}$ to rise.  \textcolor{red}{At fixed injected fraction, richer clusters are more sensitive rather than less sensitive because the number of stars in the anomalous velocity bin scales with $N_s$, while the Poisson uncertainty scales only as $\sqrt{N_s}$; the chi-square contributions in Equation~\ref{eq:Qs_def} therefore grow as the feature becomes better resolved.} The rate at which $C$ increases with $f$ depends strongly on $N_{\rm core}$: in the very well--sampled cluster NGC~6121 ($N_{\rm core}\simeq 1.6\times 10^4$), $C_{\rm ctrl}$ crosses the Tier~1 threshold $C_{\rm high}=2$ already by $f\sim 0.16$, whereas in lower--$N$ systems such as IC~1276, a much larger crystal fraction is required to produce a comparable increase.

A striking feature of Figure~\ref{fig:C_vs_f_six} is that, in several hosts, $C$ does not grow indefinitely with $f$: once $f$ approaches unity, both $C_{\rm all}$ and especially $C_{\rm ctrl}$ drop back towards values of order unity, and in some cases nearly to zero.  This \emph{does not} mean that a nearly pure crystal would be dynamically unstructured.  Instead, it is a consequence of how $Q_{\rm vel,local}$ is defined.  In each radial shell we first estimate a robust dispersion $\sigma_s$ from the \emph{current} velocities and then work in normalized units $x = v_{\rm tan}/\sigma_s$, comparing the resulting histogram to a unit--scale Rayleigh distribution.  When only a modest fraction of stars belong to the ultra--cold component, $\sigma_s$ is still set mainly by the hot background and the cold spike appears as a strong deviation from the Rayleigh form, maximizing $Q_{\rm vel,local}$ and therefore $C$.  \textcolor{red}{The apparent peak near $f\simeq 0.7$--0.8 in several panels is the transition where the IQR-based scale estimate begins to be controlled by the injected cold component rather than by the original hot background.} When $f$ is pushed close to 1, the cold component itself determines $\sigma_s$, the entire distribution becomes very narrow, and after rescaling by this smaller $\sigma_s$ the normalized velocities again resemble a single Rayleigh--like distribution.  In other words, our statistic is most sensitive to \emph{mixtures} of hot and cold populations; once the crystal completely dominates the shell, the velocity field becomes effectively single--component and $Q_{\rm vel,local}$ decreases.

This behaviour has two important implications.  First, for the range of $f$ that we consider relevant for realistic substructures ($f\lesssim 0.6$ in shell~2), the metric behaves monotonically in the sense that larger cold fractions produce larger $C$, and hence the injected crystals are clearly detected whenever they involve a sufficiently large fraction of the shell--2 stars.  Second, the drop in $C$ at very high $f$ reminds us that our present metric is not optimized for the extreme scenario of an entire shell (or entire cluster) being rearranged into a single ultra--cold component.  Our injection tests and the resulting limits should therefore be interpreted as constraints on \emph{embedded} cold components in an otherwise ''normal'' background, not as a fully general search for arbitrarily exotic phase--space configurations.

\subsection{Detection thresholds and upper limits}
\label{subsec:detection_limits}

The injection experiments described above can be summarized in terms of approximate detection thresholds on the fraction of stars that can participate in an ultra--cold component without driving the cluster into the high--$C$ regime. For each host cluster we define a 90\% detection threshold $f_{{\rm shell},90}$ as the smallest value of $f$ for which the control--ensemble crystallization index $C_{\rm ctrl}$ exceeds the Tier~1 threshold $C_{\rm high}=2$ in at least $90\%$ of the Monte Carlo realizations. Table~\ref{tab:inj_thresholds} lists $f_{{\rm shell},90}$ for each host, together with the total number of core stars $N_{\rm core}$ in the range $0.5\le R_{\rm norm}\le 3.0$.

\begin{deluxetable*}{lccc}
\tablecaption{Detection thresholds for ultra--cold shell components in Tier~4 control clusters\label{tab:inj_thresholds}}
\tablehead{
\colhead{Cluster} &
\colhead{$N_{\rm core}$} &
\colhead{$f_{{\rm shell},90}$} &
\colhead{Qualitative sensitivity} \\
\colhead{} &
\colhead{(0.5--3 $r_h$)} &
\colhead{(min.\ $f$ with $P_{\rm det,ctrl}\ge 0.9$)} &
\colhead{}
}
\startdata
Ter~3      &   770   & Never reach Tier~1 & very weak (no 90\% detection) \\
IC~1276    &  1264   & 0.60    & weak (low $N_{\rm core}$)     \\
NGC~6093   &  2552   & 0.30    & moderate                      \\
NGC~6838   &  3589   & 0.36    & moderate                      \\
NGC~4833   &  6906   & 0.19    & strong (well--sampled)        \\
NGC~6121   & 15886   & 0.16    & very strong (rich core sample)\\
\enddata
\tablecomments{
Detection thresholds for ultra--cold, single--shell tangential--velocity components injected into Tier~4 control clusters. Here $N_{\rm core}$ is the number of high--probability members in the normalized radial range $0.5 \le R_{\rm norm} \le 3.0$, and $f_{{\rm shell},90}$ is the minimum injected fraction of stars in the shell $1 \le R_{\rm norm} < 2$ for which the control--calibrated crystallization index $C_{\rm ctrl}$ exceeds the Tier~1 threshold ($C_{\rm ctrl} \ge C_{\rm high}=2$) in at least $90\%$ of Monte Carlo realizations. For Ter~3, $P_{\rm det,ctrl}$ never reaches 0.9 over the explored range of $f$, so $f_{{\rm shell},90}$ is formally undefined.}
\end{deluxetable*}

\begin{deluxetable*}{lccccc}
\tablecaption{Cluster Crystallization Statistics I\label{tab:tiers1}}
\tablehead{
    \colhead{Cluster} &
    \colhead{$N_{\rm stars}$} &
    \colhead{$z_{\rm rad}$} &
    \colhead{$z_{\rm vel}$} &
    \colhead{$C_{\rm index}$} &
    \colhead{Tiers}
}
\startdata
 NGC\_104 & $72541$ & $1.44$ & $2.36$ & $2.76$ & $1$ \\
    NGC\_5139 & $115253$ & $1.37$ & $2.27$ & $2.65$ & $1$ \\
    BH\_140 & $2727$ & $2.39$ & $-0.09$ & $2.39$ & $1$ \\
    NGC\_6205 & $21743$ & $1.01$ & $1.70$ & $1.97$ & $2$ \\
    NGC\_6752 & $28432$ & $0.46$ & $1.75$ & $1.81$ & $2$ \\
    NGC\_6496 & $1420$ & $1.63$ & $-0.37$ & $1.63$ & $2$ \\
    NGC\_3201 & $19497$ & $0.58$ & $1.47$ & $1.58$ & $2$ \\
    NGC\_6809 & $18170$ & $0.30$ & $1.54$ & $1.57$ & $2$ \\
    NGC\_1904 & $4418$ & $1.42$ & $0.49$ & $1.50$ & $2$ \\
    NGC\_5904 & $15947$ & $-0.21$ & $1.34$ & $1.34$ & $2$ \\
    NGC\_6656 & $27137$ & $0.64$ & $1.11$ & $1.28$ & $2$ \\
    NGC\_6397 & $22998$ & $0.40$ & $1.15$ & $1.22$ & $2$ \\
    NGC\_6266 & $2940$ & $1.20$ & $-2.01$ & $1.20$ & $2$ \\
    NGC\_5272 & $13196$ & $-0.25$ & $1.15$ & $1.15$ & $2$ \\
    NGC\_6254 & $16228$ & $-1.03$ & $1.15$ & $1.15$ & $2$ \\
    NGC\_6293 & $839$ & $1.10$ & $-0.99$ & $1.10$ & $2$ \\
    NGC\_6637 & $1337$ & $1.05$ & $-0.62$ & $1.05$ & $2$ \\
    NGC\_288 & $7181$ & $0.61$ & $0.56$ & $0.83$ & $3$ \\
    NGC\_6362 & $10650$ & $-0.27$ & $0.80$ & $0.80$ & $3$ \\
    IC\_4499 & $1084$ & $0.75$ & $0.23$ & $0.78$ & $3$ \\
    NGC\_6366 & $3934$ & $0.76$ & $-0.44$ & $0.76$ & $3$ \\
    NGC\_5024 & $6080$ & $-0.59$ & $0.75$ & $0.75$ & $3$ \\
    NGC\_6341 & $8988$ & $0.16$ & $0.71$ & $0.73$ & $3$ \\
    NGC\_4372 & $13224$ & $-1.86$ & $0.71$ & $0.71$ & $3$ \\
    NGC\_6441 & $916$ & $0.70$ & $-2.76$ & $0.70$ & $3$ \\
    NGC\_6218 & $12727$ & $-0.26$ & $0.66$ & $0.66$ & $3$ \\
    NGC\_6723 & $5983$ & $0.62$ & $0.15$ & $0.63$ & $3$ \\
    NGC\_1851 & $4819$ & $-1.44$ & $0.63$ & $0.63$ & $3$ \\
    NGC\_7089 & $5219$ & $-0.51$ & $0.57$ & $0.57$ & $3$ \\
    NGC\_6544 & $6705$ & $0.55$ & $-0.71$ & $0.55$ & $3$ \\
    NGC\_2808 & $7499$ & $-0.05$ & $0.55$ & $0.55$ & $3$ \\
    NGC\_6388 & $1687$ & $0.54$ & $-2.38$ & $0.54$ & $3$ \\
    NGC\_362 & $7792$ & $-1.20$ & $0.54$ & $0.54$ & $3$ \\
    NGC\_6779 & $4494$ & $0.28$ & $0.46$ & $0.53$ & $3$ \\
    NGC\_7099 & $5828$ & $-1.39$ & $0.53$ & $0.53$ & $3$ \\
    NGC\_6584 & $2021$ & $0.53$ & $-0.13$ & $0.53$ & $3$ \\
    NGC\_6101 & $4434$ & $-1.51$ & $0.50$ & $0.50$ & $3$ \\
    NGC\_4590 & $5250$ & $-0.42$ & $0.49$ & $0.49$ & $3$ \\
    NGC\_6402 & $5905$ & $-0.23$ & $0.49$ & $0.49$ & $3$ \\
    NGC\_1261 & $3124$ & $-0.51$ & $0.46$ & $0.46$ & $3$ \\
    NGC\_5897 & $4519$ & $-1.52$ & $0.42$ & $0.42$ & $3$ \\
    NGC\_6235 & $923$ & $0.37$ & $-0.35$ & $0.37$ & $3$ \\
    NGC\_7078 & $5766$ & $-0.88$ & $0.36$ & $0.36$ & $3$ \\
    NGC\_5286 & $3879$ & $-1.03$ & $0.36$ & $0.36$ & $3$ \\
    NGC\_5634 & $867$ & $-0.37$ & $0.34$ & $0.34$ & $3$ \\
    NGC\_6171 & $5687$ & $-1.17$ & $0.27$ & $0.27$ & $3$ \\
    NGC\_6981 & $2070$ & $-0.56$ & $0.18$ & $0.18$ & $3$ \\
\enddata
\end{deluxetable*}

\begin{deluxetable*}{lccccc}
\tablecaption{Cluster Crystallization Statistics II\label{tab:tiers2}}
\tablehead{
    \colhead{Cluster} &
    \colhead{$N_{\rm stars}$} &
    \colhead{$z_{\rm rad}$} &
    \colhead{$z_{\rm vel}$} &
    \colhead{$C_{\rm index}$} &
    \colhead{Tiers}
}
\startdata
            NGC\_5466 & $2935$ & $-2.03$ & $0.15$ & $0.15$ & $3$ \\
    NGC\_6864 & $990$ & $-0.12$ & $0.14$ & $0.14$ & $3$ \\
    NGC\_5986 & $3111$ & $-0.50$ & $0.11$ & $0.11$ & $3$ \\
    NGC\_6541 & $6790$ & $-1.29$ & $0.10$ & $0.10$ & $3$ \\
        NGC\_6139 & $1153$ & $0.07$ & $-0.85$ & $0.07$ & $3$ \\
    NGC\_6229 & $719$ & $-1.01$ & $0.07$ & $0.07$ & $3$ \\
    NGC\_6934 & $1902$ & $-0.85$ & $0.05$ & $0.05$ & $3$ \\
    NGC\_2298 & $2188$ & $-0.60$ & $0.03$ & $0.03$ & $3$ \\
    Ter\_3 & $770$ & $-0.96$ & $-1.29$ & $0.00$ & $4$ \\
    NGC\_4833 & $6906$ & $-1.87$ & $-0.25$ & $0.00$ & $4$ \\
    NGC\_6287 & $1576$ & $-0.03$ & $-0.38$ & $0.00$ & $4$ \\
    NGC\_6333 & $1828$ & $-0.77$ & $-1.37$ & $0.00$ & $4$ \\
    NGC\_6144 & $2018$ & $-0.68$ & $-0.06$ & $0.00$ & $4$ \\
    NGC\_6352 & $2828$ & $-1.23$ & $-1.61$ & $0.00$ & $4$ \\
    NGC\_6121 & $15886$ & $-0.69$ & $-0.66$ & $0.00$ & $4$ \\
    NGC\_6093 & $2552$ & $-1.20$ & $-0.13$ & $0.00$ & $4$ \\
    NGC\_5927 & $1820$ & $-0.23$ & $-1.80$ & $0.00$ & $4$ \\
    NGC\_5824 & $912$ & $-1.05$ & $-0.48$ & $0.00$ & $4$ \\
    NGC\_6539 & $1175$ & $-0.69$ & $-0.42$ & $0.00$ & $4$ \\
    NGC\_6553 & $1063$ & $-0.12$ & $-2.49$ & $0.00$ & $4$ \\
    NGC\_5053 & $1635$ & $-0.49$ & $-0.06$ & $0.00$ & $4$ \\
    NGC\_6652 & $987$ & $-0.89$ & $-0.76$ & $0.00$ & $4$ \\
    NGC\_6626 & $2375$ & $-1.25$ & $-1.74$ & $0.00$ & $4$ \\
    NGC\_6681 & $1940$ & $-0.80$ & $-0.43$ & $0.00$ & $4$ \\
    NGC\_2419 & $928$ & $-0.16$ & $-0.38$ & $0.00$ & $4$ \\
    FSR\_1758 & $991$ & $-0.52$ & $-0.76$ & $0.00$ & $4$ \\
    NGC\_6838 & $3589$ & $-0.50$ & $-1.01$ & $0.00$ & $4$ \\
    NGC\_6712 & $1867$ & $-0.47$ & $-0.38$ & $0.00$ & $4$ \\
    IC\_1276 & $1264$ & $-0.73$ & $-1.06$ & $0.00$ & $4$ \\
    NGC\_4147 & $772$ & $-0.74$ & $-0.26$ & $0.00$ & $4$ \\
    NGC\_6715 & $1749$ & $-1.62$ & $-0.35$ & $0.00$ & $4$ \\
    NGC\_6273 & $5821$ & $-2.16$ & $-0.04$ & $0.00$ & $4$ \\
\enddata
\tablecomments{
Table \ref{tab:tiers1} and \ref{tab:tiers2} together form the Cluster statistics and crystallization tiers of the full 79 sample. 
$N_{\rm stars}$ is the number of high--probability ($P_{\rm mem}\ge 0.9$) members with 
$0.5 \le R_{\rm norm} \le 3.0$ that enter the structure metrics. 
$z_{\rm rad}$ and $z_{\rm vel}$ are the standardized radial and kinematic components defined in 
Section~\ref{sec:metrics}, and $C_{\rm index}$ is the combined crystallization index from 
Equation~(\ref{eq:Cindex_def}). The ''Tiers'' column encodes the four--level classification adopted in 
Section~\ref{subsec:tier_pop}: Tier~1 (high--$C$) clusters have $C_{\rm index}\ge C_{\rm high}=2$; 
Tier~2 (intermediate) clusters have $C_{\rm mid}\le C_{\rm index}<C_{\rm high}$ with $C_{\rm mid}=1$; 
Tier~3 (low) clusters have $0\le C_{\rm index}<C_{\rm mid}$; and Tier~4 clusters have $C_{\rm index}=0$ 
and define the control population used in the injection experiments of Section~\ref{sec:injections}.}
\end{deluxetable*}

\clearpage

Taken together, these results imply that in the best--sampled control clusters, NGC~6121 and NGC~4833, our crystallization metric would be sensitive to ultra--cold, shell--confined components involving only $\simeq 0.2$ of the shell--2 stars ($f_{{\rm shell},90}\approx 0.16$--0.19). Since shell~2 typically contains roughly half of the stars in the $0.5\le R_{\rm norm}\le 3.0$ core sample for these systems, this corresponds to detecting cold structures that comprise only a few to several percent of all core stars. Intermediate--$N$ hosts such as NGC~6093 and NGC~6838 require somewhat larger injected fractions ($f_{{\rm shell},90}\approx 0.3$--0.36), while in the low--$N$ host IC~1276 we achieve 90\% detection probability only when the crystal dominates shell~2 ($f_{{\rm shell},90}\simeq 0.6$). In the extreme low--$N$ case of Ter~3, even an injection with $f=1$ fails to produce $P_{\rm det,ctrl}\ge 0.9$, underscoring the limited constraining power of very sparsely sampled clusters.

Because these six Tier~4 hosts span $N_{\rm core}$ from $\sim 8\times 10^2$ to nearly $1.6\times 10^4$ and cover a range of structural properties, they provide a reasonable indication of the sensitivity of our method across the broader control population. In the absence of high crystallization signals in the Tier~4 sample, we can therefore state that no ultra--cold, single--shell tangential--velocity components of the type modelled here, involving more than at most a few to $\sim 10$--20\% of the core stars (depending on $N_{\rm core}$), are present in the \emph{Gaia}--accessible core regions of these dynamically smooth globular clusters.

\section{Discussion}
\label{sec:discussion}

\subsection{Interpretation of the crystallization index}

Our crystallization framework is deliberately model--light: rather than fitting detailed dynamical models to each cluster, we use simple, standardized statistics to quantify how far a system deviates from a smooth, quasi--equilibrium baseline. The two ingredients, $Q_{\rm rad}$ and $Q_{\rm vel,local}$ (Section~\ref{sec:metrics}), probe complementary aspects of projected phase--space structure.

The radial metric $Q_{\rm rad}$ asks whether the normalized projected radius distribution $R_{\rm norm}$ in the range $0.5 \le R_{\rm norm} \le 3$ can be described, up to noise, by a smooth one--dimensional profile. By construction, $Q_{\rm rad}$ is insensitive to the overall scale of the profile (which is removed by using $R_{\rm norm} = R_{\rm arcmin}/r_{h,{\rm arcmin}}$) and to Poisson fluctuations at very low counts (mitigated by the moving--average smoothing in Equation~\ref{eq:Q_rad_def}). Large $Q_{\rm rad}$ values correspond to significant excesses or deficits in one or more radial bins relative to a smoothed reference profile ''learned'' from the cluster's own data. Physically, this can arise from tidal distortions, shells, clumps, or strong departures from a single--scale profile.

The kinematic metric $Q_{\rm vel,local}$ asks a different question: in a given radial shell, is the one--dimensional tangential--velocity distribution consistent with the Rayleigh law expected for the norm of a two--dimensional Gaussian velocity field? Here the normalization by a shell--level robust dispersion $\sigma_s$ removes the overall scale of the velocity field, so that only the \emph{shape} of the distribution matters. Large $Q_{\rm vel,local}$ values mean that, after this normalization, the histogram of $x=v_{\rm tan}/\sigma_s$ deviates strongly from the Rayleigh reference in a chi--square sense (Section~\ref{subsec:Qvellocal_def}), typically because of pronounced wings, multi--modality, or an overabundance of very slow or very fast stars.

By converting both raw metrics to standardized $z$--scores and combining them into
\begin{equation}
  C_{\rm index} \equiv \sqrt{z_{\rm rad,pos}^2 + z_{\rm vel,pos}^2}\,,
\end{equation}
we effectively embed each cluster in a two--dimensional space of ''how radially lumpy'' and ''how kinematically odd'' it is compared to the population as a whole. The fact that known dynamically complex systems such as $\omega$~Cen and 47~Tuc rank highly in $C_{\rm index}$, while many ordinary clusters lie near $C_{\rm index}\approx 0$, suggests that the metric is successfully capturing real physical diversity in dynamical states, without being tied to any specific microphysical scenario.

\subsection{Relation to classical views of globular cluster structure}

\textcolor{red}{There is a large literature on the structural and dynamical complexity of Galactic globular clusters, including multiple stellar populations, internal rotation, anisotropy, higher-order proper-motion moments, and tidal features \citep[e.g.][]{Gratton2012,MiloneMarino2022,Kamann2018,Heyl2017_47Tuc,Ziliotto2025_47TucKinematics,Ziliotto2026OmegaCen}. In that context, our crystallization index should be viewed as a minimalist ensemble-screening statistic, not a replacement for population-resolved kinematic modelling. Moment-based diagnostics such as skewness and kurtosis measure the shape of the proper-motion component distributions directly, often as functions of stellar population and radius. Our $Q_{\rm vel,local}$ instead compresses the radial-shell distribution of cluster-centric speed magnitudes into a Rayleigh-deviation score, while $Q_{\rm rad}$ adds a separate spatial term.}

\textcolor{red}{The connection is nevertheless physically meaningful. NGC~104 and NGC~5139, two of our high-$C_{\rm index}$ clusters, are also systems where independent studies find strong non-Gaussian or population-dependent proper-motion structure. This agreement suggests that $z_{\rm vel}$ is sensitive to at least some of the same higher-order kinematic complexity captured by skewness/kurtosis analyses, even though it does not preserve the full vector information. Conversely, BH~140 illustrates the complementarity of the radial channel: it is high-$C$ because of $z_{\rm rad}$ rather than $z_{\rm vel}$, and would not be highlighted by a velocity-moment diagnostic alone.}

\textcolor{red}{Physically, high $C_{\rm index}$ can be produced by several standard mechanisms. Multiple stellar populations can have different radial concentrations and velocity anisotropies; rotation and perspective effects can distort the local speed distribution; tidal stripping, disk or bulge shocks, and eccentric orbits can create radial asymmetries or cold debris; incomplete phase mixing after early mergers or accretion can preserve coherent substructure; and mass segregation, binaries, or retained remnants can reshape local velocity distributions. Catalogue effects, especially crowding-dependent incompleteness or residual field contamination, also remain possible for individual clusters. The metric is therefore best used as a triage tool for deciding where more detailed chemodynamical modelling is warranted.}

\textcolor{red}{The correlation analysis in Section~\ref{subsec:C_vs_env} supports this interpretation. Raw $C_{\rm index}$ depends strongly on $N_{\rm core}$, weakly on mass, and modestly on a relaxation-time proxy, while $C_{\rm resid}$ reduces the mass and sample-size trends but retains substantial scatter. Crystallization therefore does not reduce to a simple function of one global parameter.}

\textcolor{red}{Our approach is complementary to more detailed dynamical modelling, such as Jeans or $N$--body fits to rotation fields, anisotropy profiles, higher--order moments, and multi-epoch astrometric catalogues \citep[e.g.][]{Kamann2018,Libralato2022HACKS}. Those methods attempt to reconstruct a physically motivated phase--space distribution function for each cluster individually. By contrast, $C_{\rm index}$ treats each cluster as a point in a low--dimensional summary space, making it easier to perform ensemble--level comparisons and to identify outliers that deserve more detailed follow--up.}

\subsection{Implications for globular clusters as technosignature targets}

Several authors have argued that globular clusters may be promising environments for long--lived technological civilizations, owing to their old ages, long dynamical lifetimes, and small typical separations between stars \citep{DiStefanoRay2016}. In parallel, this prospect has recently motivated an increasing number of dedicated technosignature surveys targeting globular clusters\citep{huang2025fast}. In such a scenario, the phase--space structure of the cluster becomes directly relevant to questions of habitability and interstellar engineering: the local velocity dispersion, tidal field, and long--term stability of orbits all influence the ease of travel, infrastructure building, and communication within the cluster.

\textcolor{red}{The low metallicities of many GCs are an important caveat for any habitability or technosignature interpretation. Metal-poor environments may suppress the formation of giant planets, and early searches for short-period planets in 47~Tuc found no detections at the surveyed sensitivity \citep{Gilliland2000_47TucPlanets}. On the other hand, the dependence of small-planet occurrence on metallicity is weaker than that of giant planets \citep{Buchhave2012SmallPlanetsMetallicity}, and the dense, old nature of GCs motivates technosignature searches for reasons that are not identical to ordinary exoplanet yield optimization. We therefore present $C_{\rm index}$ only as a dynamical target-prioritization layer, not as a claim that high- or low-$C$ clusters are intrinsically more habitable.}

Our crystallization index does not attempt to model technosignatures explicitly, but it does provide a quantitative handle on how dynamically ''ordinary'' or ''exceptional'' a given cluster is, relative to the Milky Way GC population. One can imagine at least two broad, complementary ways to use this information in a SETI context:

\begin{itemize}[leftmargin=*]
\item Conservative strategy: target smooth, dynamically ordinary clusters. 
  If one hypothesizes that long--lived, distributed technological activity is most likely to arise in clusters that are dynamically stable and free from extreme tidal or merger events, then low--$C_{\rm index}$ clusters become attractive targets: they are old, relaxed systems whose present--day phase--space structure closely matches simple equilibrium expectations. In this view, Tier~4 clusters define a baseline for ''quiet'' environments in which technosignatures, if present, might be driven more by internal social and technological processes than by external dynamical perturbations.
\item Exploratory strategy: target dynamically exceptional clusters. 
  Alternatively, one may speculate that clusters with unusual dynamical histories, those with large $C_{\rm index}$, could be special in ways that are also relevant for the emergence or survival of technology. For example, extreme retention of remnants, merger histories, or strong internal rotation might correlate with atypical planet formation environments, radiation fields, or long--term stability of habitable zones. In this speculative picture, Tier~1 clusters would be natural laboratories to search for unconventional technosignatures or large--scale engineering projects.
\end{itemize}

\subsection{Limitations and avenues for improvement}

\textcolor{red}{Several limitations of the current study are worth highlighting. First, our analysis uses Gaia-based proper motions and parallaxes to construct plane-of-sky, cluster-centric tangential velocities; line-of-sight velocities are only implicit in the membership and cluster-parameter catalogues. A full 3D treatment, combining Gaia with large spectroscopic surveys and integral-field spectroscopy \citep[e.g.][]{Kamann2018}, would allow more powerful anisotropy and phase-space diagnostics. Second, related astrometric resources such as the HST Astrometry Catalog of Galactic Globular Clusters \citep{Libralato2022HACKS} and population-resolved proper-motion studies \citep[e.g.][]{Ziliotto2025_47TucKinematics,Ziliotto2026OmegaCen} already explore several of these directions for selected clusters; our contribution is the homogeneous ensemble-level scalar rather than the definitive kinematic model of any one cluster. Third, we normalize radii by the projected half-light radius $r_h$, which is convenient but may blur information about multi-scale structures such as distinct cores, envelopes, and extended tidal features. Alternative normalizations or multi-scale versions of $Q_{\rm rad}$ could be explored. Fourth, the null models for $Q_{\rm rad}$ and $Q_{\rm vel,local}$ are intentionally simple. Real clusters may deviate from these baselines in ways that are entirely consistent with standard dynamics. Finally, our injected ``crystals'' are extremely narrow in normalized velocity and confined to a single shell. The limits in Section~\ref{sec:injections} therefore apply only to embedded ultra-cold single-shell components of the specified form, not to all possible exotic or natural substructures.}

Despite these limitations, the present crystallization analysis demonstrates that simple, ensemble--normalized metrics can reveal meaningful structure in the GC population and can be pushed to the point of providing quantitative non--detections of certain classes of cold substructure. This suggests several natural extensions: coupling $C_{\rm index}$ to multi--population maps \citep[e.g.][]{Gratton2012,MiloneMarino2022}, incorporating 3D velocities and orbits, and embedding the framework in forward models that connect hypothetical engineered phase--space structures to observable kinematic signatures.

\section{Conclusions}
\label{sec:conclusions}

We have introduced a simple, \emph{Gaia}--anchored framework for quantifying ``kinematic crystallization'' in Galactic globular clusters. Starting from the homogeneous catalogues of \citet{BaumgardtVasiliev2021,VasilievBaumgardt2021}, we constructed a working sample of 79 clusters with well--measured projected radii and tangential velocities (Section~\ref{sec:data}). Two model--light statistics summarize their small--scale phase--space structure: a radial inhomogeneity metric $Q_{\rm rad}$ that measures bin--to--bin lumpiness in the normalized radius distribution $R_{\rm norm}$, and a local tangential--velocity metric $Q_{\rm vel,local}$ that tests for deviations from the Rayleigh form expected for an isotropic 2D Gaussian field (Section~\ref{sec:metrics}). After standardizing these to $z$--scores and combining their positive parts in quadrature, we obtain a crystallization index $C_{\rm index}$ that ranks clusters by the joint extremeness of their radial and kinematic structure.

Across the 79--cluster ensemble, the distribution of $(z_{\rm rad},z_{\rm vel})$ is strongly peaked near $(0,0)$, with most systems having $C_{\rm index}<1$ and therefore lying in our Tier~2--4 regime. Only three clusters exceed the Tier~1 threshold $C_{\rm high}=2$, and even these lie within the broad expectations of an analytic smooth--null model: approximating $z_{\rm rad}$ and $z_{\rm vel}$ as independent standard normal variates, one expects $4.5\pm 2.0$ high--$C$ clusters out of 79, and the probability of observing at least three is $\simeq 0.83$ (Section~\ref{subsec:global_C}). Thus the high--$C$ tail is not itself a statistically significant excess over a smooth baseline, but the Tier~1 clusters remain individually interesting as dynamically complex outliers. A four--tier classification based on $C_{\rm index}$ (Tier~1: $C\ge 2$; Tier~2: $1\le C<2$; Tier~3: $0<C<1$; Tier~4: $C=0$) provides a convenient vocabulary for describing this diversity.

\textcolor{red}{We have shown that $C_{\rm index}$ is partly sensitive to sample size and must be interpreted accordingly. The raw index correlates with $\log_{10}N_{\rm core}$ ($\rho=0.57$), while the $N_\star$--corrected index $C_{\rm resid}$ removes this trend to first order (Section~\ref{subsec:C_vs_N}). Correlations with present-day mass are weak, and a dynamical-age proxy shows the expected tendency for more relaxed clusters to have lower crystallization (Section~\ref{subsec:C_vs_env}). We do not present a metallicity correlation in this work because metallicity is not part of the homogeneous input table used by the analysis.}

To quantify what kinds of ultra--cold components our metric would detect, we performed controlled injection experiments in six low--$C$ clusters drawn from the Tier~4 pool: IC~1276, NGC~4833, NGC~6093, NGC~6121, NGC~6838, and Ter~3 (Section~\ref{sec:injections}). Into each cluster we injected synthetic, extremely cold tangential--velocity components confined to $1\le R_{\rm norm}<2$ and occupying a fraction $f$ of shell--2 stars. Using the Tier~4 population to calibrate the velocity $z$--scores, we measured the detection probability that the control--ensemble index $C_{\rm ctrl}$ exceeds the Tier~1 threshold. The resulting 90\% detection limits span $f_{{\rm shell},90}\simeq 0.16$--0.19 in the best--sampled hosts (NGC~6121 and NGC~4833), $f_{{\rm shell},90}\simeq 0.3$--0.36 in intermediate--$N$ systems (NGC~6093, NGC~6838), and $\gtrsim 0.6$ or undetected over our grid in the lowest--$N$ controls (IC~1276, Ter~3). In the well--sampled clusters this corresponds to cold components involving only a few percent of core stars, whereas in sparse systems only nearly dominant crystals would be detectable. The absence of strong crystallization signals in the Tier~3--4 sample therefore rules out ultra--cold, single--shell structures of this type at the level of $\sim$few--to--$\sim$20\% of the core stars, within the $0.5\le R_{\rm norm}\le 3.0$ region accessible to \emph{Gaia}.

Taken together, these results suggest that the present--day Milky Way globular cluster system is largely consistent with smooth, equilibrium phase--space structure at the resolution probed here, punctuated by a small number of dynamically complex outliers but lacking any ``smoking gun'' crystallization that would demand non--standard physics or artificial engineering. The crystallization index provides a compact way to rank clusters by dynamical extremeness, to select targets for deeper chemodynamical modelling, and to define a controlled, low--$C$ baseline for future injection and sensitivity studies. Extensions that incorporate full 3D velocities, explicit multiple--population information, and forward modelling of specific crystallization scenarios---natural or artificial---will further sharpen the connection between observed phase--space structure, cluster formation and evolution, and any putative technosignatures in dense stellar environments.

\begin{acknowledgments}
This work was supported by the National Key R\&D Program of China (No.\ 2024YFA1611804), 
the China Manned Space Program (CMS-CSST2025-A01), the National SKA Program of China under Grant No. 2025SKA0120104, 
the Shandong Provincial Natural Science Foundation (ZR2024QA180), 
and the Scientific Research Fund of Dezhou University (4022504019). This work has made use of data from the European Space Agency (ESA) mission
\emph{Gaia} (https://www.cosmos.esa.int/gaia), processed by the \emph{Gaia}
Data Processing and Analysis Consortium (DPAC). Funding for the DPAC has been
provided by national institutions, in particular the institutions participating
in the \emph{Gaia} Multilateral Agreement. We acknowledge the use of the public catalogue “Catalogue of stars in Milky Way globular clusters from Gaia EDR3” by E. Vasiliev and H. Baumgardt, available via Zenodo \cite{zenode}.

\end{acknowledgments}

\bibliography{sample701}{}
\bibliographystyle{aasjournalv7}

\end{document}